\documentclass{article}

\usepackage{arxiv}

\usepackage[utf8]{inputenc} 
\usepackage[T1]{fontenc}    
\usepackage{hyperref}       
\usepackage{url}            
\usepackage{booktabs}       
\usepackage{amsfonts}       
\usepackage{nicefrac}       
\usepackage{microtype}      
\usepackage{lipsum}		
\usepackage{graphicx}
\usepackage{natbib}
\usepackage{doi}
\usepackage{amsmath}
\usepackage{amssymb}
\usepackage{enumitem}
\usepackage{pdfpages}
\usepackage{blkarray}
\usepackage{mathtools}
\usepackage{natbib}
\usepackage{url}
\usepackage[lined,boxed,linesnumbered]{algorithm2e}
\usepackage{graphicx,subfigure}
\usepackage{epstopdf} 
\usepackage{array}

\newcommand{\bbeta}{\mbox{\boldmath{$\beta$}}}

\newcommand{\bgamma}{\mbox{\boldmath{${\gamma}$}}}
\newcommand{\balpha}{\mbox{\boldmath{${\alpha}$}}}
\newcommand{\btheta}{\mbox{\boldmath{${\theta}$}}}

\DeclareMathOperator*{\argmin}{arg\,min}

\def\smt{{\mbox{\tiny T}}}

\def\bY{\mathbf Y}
\def\bX{\mathbf X}
\def\bx{{\bf x}}
\def\by{{\bf y}}
\def\bz{{\bf z}}
\def\bZ{{\bf Z}}

\def\bg{\mathbf g}

\def\bA{\mathbf A}
\def\bB{\mathbf B}
\def\bU{\mathbf U}
\def\bG{\mathbf G}
\def\bW{\mathbf W}
\def\bw{\mathbf w}
\def\bX{\mathbf X}
\def\bx{\mathbf x}
\def\bI{\mathbf I}

\def\bV{{\bf V}}

\def\bD{\mathbf D}
\usepackage{color}
\usepackage{comment}
\usepackage{ulem}
\usepackage{soul}

\title{Scalable Randomized Kernel Methods for Multiview Data Integration and Prediction}

\author{ \href{https://orcid.org/0000-0001-9593-4778}{\includegraphics[scale=0.06]{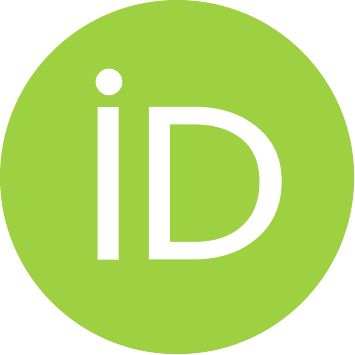}\hspace{1mm}Sandra E. Safo}\thanks{Corresponding Author: Sandra Safo, www.sandraesafo.com} \\
	Division of Biostatistics\\
	University of Minnesota Twin Cities\\
	Minneapolis, MN 55455 \\
	\texttt{ssafo@umn.edu} \\
	\And
	{\hspace{1mm}Han Lu} \\
	Division of Biostatistics\\
	University of Minnesota Twin Cities\\
	Minneapolis, MN 55455 \\
	\texttt{lu000054@umn.edu} \\
}

\renewcommand{\shorttitle}{RandMVLearn for Multiview Data Integration}

\hypersetup{
pdftitle={A template for the arxiv style},
pdfsubject={q-bio.NC, q-bio.QM},
pdfauthor={David S.~Hippocampus, Elias D.~Striatum},
pdfkeywords={First keyword, Second keyword, More},
}

\begin{document}
\maketitle

\begin{abstract}
We develop scalable randomized kernel methods for jointly associating data from multiple sources and simultaneously predicting an outcome or classifying a unit into one of two or more classes. The proposed methods model nonlinear relationships in multiview data together with predicting a clinical outcome and are capable of identifying variables or groups of variables that best contribute to the relationships among the views. We use the idea that random Fourier bases can approximate shift-invariant kernel functions to construct nonlinear mappings of each view and we use these mappings and the outcome variable to learn view-independent low-dimensional representations. Through simulation studies, we show that the proposed methods outperform several other linear and nonlinear methods for multiview data integration. When the proposed methods were applied to gene expression, metabolomics, proteomics, and lipidomics data pertaining to COVID-19, we identified several molecular signatures forCOVID-19 status and severity. 
Results from our real data application and simulations with small sample sizes suggest that the proposed methods may be useful for small sample size problems.\\
\noindent\textbf{Availability:} Our algorithms are implemented in Pytorch and interfaced in R and would be made available at:
\url{https://github.com/lasandrall/RandMVLearn}.\\

\end{abstract}


\section{Introduction}
\label{s:intro}

Many biomedical research generates multiple types of data (e.g. genomics, proteomics), measured on the same set of individuals, and a typical goal is to understand complex disease mechanisms through these unique, but complementary data. 
For many complex diseases, established approaches to treatment and risk stratification are available, but often do not address the underlying mechanisms but the symptoms. Understanding the pathobiology of complex diseases requires an analytical approach that goes beyond separate analysis of each data type.

\subsection{COVID-19 Molecular Study}
There is still much to be learned about the pathobiology of COVID-19.  Research suggests that patients with and without severe COVID-19 have different genetic, pathological, and clinical signatures \citep{geneticscovid,overmyer:2020}. This highlights a 
need to use multiple molecular data to better  understand the severity of the  disease. 
Our work is motivated by data pertaining to COVID-19 analyzed in \cite{overmyer:2020}.  Blood samples were obtained from 128 patients who had moderate to severe respiratory problems similar to COVID-19 and admitted to Albany Medical Center, NY, from 6 April 2020 to 1 May 2020.  Of these, $102$ and $26$ tested positive and negative for COVID-19, respectively. Blood samples were quantified for metabolomics, RNA sequencing (RNA-Seq), proteomics, and lipidomics. 
In \cite{overmyer:2020}, machine learning methods including lasso were used to associate molecules with disease severity and to determine key determinants of COVID-19 severity. The primary analyses focused on individual investigation of each molecular data. Pairwise correlation analysis was conducted to associate pairs of molecular data in a secondary analysis.

We take a holistic approach to simultaneously integrate the outcome and the multiple molecular data, thus leveraging information across all molecular data, while accounting for prior biological information. Our goal is to identify molecular signatures and pathways that can contribute to the severity and status of COVID-19, ultimately shedding more light into the pathogenesis of COVID-19 severity. Two methods were used to measure disease severity in \cite{overmyer:2020}: i) the World Health Organization (WHO) 0-8 disease specific scale,  where 8 denotes death, and ii) a score out of 45 that indicates the number of hospital free days (HFD-45). Of note, a HFD-45 value of 0 implies  the individual was still admitted in the hospital after 45 days, or that the individual died. The HFD-45 measurement is recommended, as it is not specific to COVID-19 \citep{overmyer:2020}.  

\subsection{Existing Methods}
Many linear and nonlinear methods have been proposed for integrating data from multiple sources. Some methods learn low-dimensional representations of each view that maximally correlate the views (e.g. canonical correlation analysis [CCA \citep{Hotelling:1936}], deep CCA \citep{andrew2013deep}, kernel CCA \citep{lopez2014randomized}). Other methods learn view-independent low-dimensional representations of all the views (e.g. generalized CCA \citep{horst1961generalized,kettenring1971canonical}, deep generalized CCA [Deep GCCA \citep{benton2017deep}). Others learn both common and view-dependent low-dimensional structure in the views (e.g. JIVE \citep{lock2013joint}). These low-dimensional representations are then associated with an outcome, for downstream tasks, that may be supervised or unsupervised. We refer to such methods as two-steps. 

Methods have also been proposed that simultaneously associate the views and predict an outcome (e.g. SIDA \citep{SIDA:2019}, BIP \citep{chekouo2020bayesian}, sJIVE \citep{PALZER2022107547}, Deep IDA \citep{wang2021deep} and MOMA \citep{MOMA:2022}. We refer to these methods as one-step. These methods are different from two-step methods in that the problem of associating views and predicting an outcome is combined. In this way, the outcome variable is used to guide learning about the low-dimensional representations in the data,
likely resulting  in a more clinically meaningful findings. In this article, we develop methods for jointly associating data from multiple sources and predicting an outcome.

Most existing methods for jointly associating multiple views and predicting an outcome have focused on learning linear relationships among the views and between the views and an outcome.  However, the intrinsic relationships between the multiple views and an outcome are too complex to be understood only by linear methods. On the other hand, the limited nonlinear methods available either lack ability to provide interpretable results, or only apply to categorical outcomes, or do not utilize prior biological information or are computationally prohibitive for large-scale data \citep{hu2019multi, wang2021deep}.  
Recently, a data integration and classification method (MOMA) for multiview learning that uses attention mechanism for interpretability have been proposed \citep{MOMA:2022}. MOMA creates a module (e.g., gene set) for each view and uses attention mechanisms to identify modules and variables  related to the classification task. Notably, the detection of modules in MOMA is data driven. 
The methods mentioned above can be used for categorical outcomes but not for continuous outcomes. These methods do not permit the use of prior biological knowledge such as variable group information. Further, the resampling approach used in Deep IDA for variable ranking tends to be computationally expensive for large data.

\subsection{Our Approach}
We bring four major contributions to integrative analysis. First, we develop a scalable kernel method for learning view-independent nonlinear low-dimensional representations in multiview data that  maximize relationships among views and can predict an outcome. For computational savings, we approximate the kernel evaluations of the input data of each view by Fourier random features \citep{rahimi2008random,lopez2014randomized}. The method helps to reveal nonlinear patterns in the views and can scale to large training sizes. Second, we use an outcome to guide learning about variables that are relevant for  constructing view-independent representations. Third, to aid in interpretability, we learn variables driving the shared low-dimensional nonlinear structure in the data through the randomized nonlinear mappings. We extend the method to scenarios where  prior biological information about variables (i.e. variable groups) are available, which could yield more interpretable findings. To our knowledge, this is one of the first nonlinear-based methods for integrative analysis and prediction that does so. Fourth, we provide an efficient implementation of the proposed methods in Pytorch, and interface it with R to increase the reach of  our methods.  The results of our real data applications and simulations with small sample sizes show that the proposed methods can be useful for problems with small sample sizes.

The rest of the article is organized as follows. In Section 2, we introduce the proposed method. In Section 3, we provide our optimization procedure and algorithms for the proposed methods. In Section 4, we evaluate the prediction and variable selection accuracy of  our method using extensive simulations. In Section 5, we apply our method to the motivating data to identify molecular signatures for COVID-19 severity and status. We end with some brief discussion in Section 6.

\section{Method}
\label{s:method}

\noindent We use the following notation for the available data. For one clinical outcome, we denote it by $y_{i}$ for subject $i=1,\ldots,n$. For all subjects, we let $\by$ denote the outcome.  This can be either continuous (e.g. COVID-19 severity) or binary (e.g. COVID-19 status). If there are multiple (or $q$) continuous clinical outcomes for subject $i$, we collect these in a  $q$-length vector $\mathbf{y}_i$. For all subjects, we collect these vectors into a matrix $\bY \in \Re^{n \times q}$. Suppose that molecular (or omics) and/or phenotypic data are available from  $d=1,\ldots,D, D\ge 2$ different sources and each view is arranged in an $n$ by $p^d$ matrix $\mathcal{\mathbf{X}}^d$, where the superscript $d$ corresponds to the $d$th source. For instance, for the same set of $n$ individuals, matrix $\mathcal{\mathbf{X}}^1$ consists of metabolomics data,  $\mathcal{\mathbf{X}}^2$ consists of gene expression levels, $\mathcal{\mathbf{X}}^3$ consists of protein  levels, and $\mathcal{\mathbf{X}}^4$ consists of phenotypic data.  The phenotypic data  may contain mixed data types: continuous, categorical. We use indicator matrices to represent categorical variables \citep{gifi1990nonlinear}. We have three many goals. First, we wish to model complex nonlinear relationships among these different data types via a joint or shared low-dimensional nonlinear embedding. Second, we want to identify individual or groups of variables potentially contributing to this nonlinear relationships between the views. Third, we want to use a clinical outcome to guide the extraction of the shared  low-dimensional nonlinear representations. Joint modeling of the relationship between views and  prediction of a clinical outcome  endows the shared nonlinear representations and extracted variables or variable groups with predictive capacity and can increase interpretability.  We emphasize that these three goals will be achieved jointly. 

\subsection{Associating multiview Data}
Many existing methods for associating data from multiple views assume that there is  a common (view-independent) low-dimensional embedding, say $\bG  \in \Re^{n \times r}$ that drives the relationships among the views.  Each view is expressed as a linear function of this shared embedding, i.e.  $\bX^{d} \approx \bG \mathbf{B}^{d^{\smt}}$. 
Here $\mathbf{B}^{d}$ is a $p^d$ by $r$ loading matrix for view $d$ with each row corresponding to $r$ coefficients for a specific variable. Typically, $\mathbf{B}^{d^\smt}\mathbf{B}^d = \mathbf{I}_{r \times r}$. Thus, one can write $\mathbf{X}^d\mathbf{B}^d  \approx \mathbf{G}$. Let $\bg_i \in \Re^r$, $i=1,\ldots,n$ be the $i$th row in $\bG$. The above approximation assumes that there is a low-dimensional space $\Re^{r}$ such that subject $i$ is represented as $\mathbf{g}_i \in \Re^{r}$ in this space. For each $d=1,\ldots,D$, $\mathbf{B}^{d}$ maps this low-dimensional representation $\mathbf{g}_i$ for subject $i$ to the observation $\bx^d_i$ for that subject. These mappings have been restricted to be linear. We reformulate these mappings using nonlinear functions to capture complex nonlinear relationships in the views. 

We assume that the shared low-dimensional representation $\bg_i$ for subject $i$   is a nonlinear function of $\bx^d_i$, i.e. $\bg_i \approx \mathcal{\textbf{f}}^d(\bx^d_i), d=1,\ldots,D$, $i=1,\ldots,n$, and $\textbf{f}^d = (f_1^d,\ldots,f_r^d)$. One of our tasks is to learn the unknown functional relationship between the $p^d-$dimensional input space $\Re^{p^d}$  and the $r$-dimensional output space $\Re^{r}$. We assume that the function $\mathcal{\textbf{f}}^d$ belongs to a reproducing kernel Hilbert space (RKHS) of \textit{vector-valued} functions. 
Let $\mathcal{H}(\mathcal{\mathbf{K}}^d)$ be a vector-valued RKHS of functions $\mathbf{f}^d: \Re^{p^d} \rightarrow \Re^r$
that is defined by the positive semi-definite reproducing kernel matrix $\mathbf{K}^d$. Thus, $\mathbf{K}^d$ takes  two inputs  $\bx^d$, $\widetilde{\bx}^{d} \in \Re^{p^d}$ and produces a matrix $\Re^{r \times r}$. For a vector-valued RKHS of functions $\mathbf{f}^d: \Re^{p^d} \rightarrow \Re^r$, we have that for every $\mathbf{a}^d \in \Re^r$, and $\mathbf{x} \in \Re^{p^d}$, the product of the kernel evaluated  at $\widetilde{\mathbf{x}} \in \Re^{p^d}$, and $\mathbf{a}^d$ belongs to $\mathcal{H}$ i.e. $\mathbf{K}^d(\mathbf{x} , \mathbf{\tilde{x}})\mathbf{a} \in \mathcal{H}$. Please refer to \cite{alvarez2011kernels} and the references therein for a review of vector-valued RKHS. By the reproducing property of $\mathbf{K}^d$, for every $\mathbf{a}^d\in \Re^{r}$, and $\mathbf{x}^d\in \Re^{p^d}$, $\langle\ \mathbf{f}^d,\mathbf{K}^d(\cdot,\mathbf{x}^d)\mathbf{a}^d \rangle_{\mathbf{K}} = \mathbf{f}^{d^\smt}\mathbf{a}^d$, where $\langle\ \cdot ~\rangle_{\mathbf{K}}$ is the inner product in $\mathcal{H}$. 

For a fixed $\mathbf{G}$, we follow the standard theory of kernel learning for vector-valued output (i.e. $\bg_i \in \Re^{r}$) to define a regularization approach (e.g. \cite{alvarez2011kernels,baldassarre2012multi}) for associating the multiple views. In particular, we minimize the regularized empirical error: 
\begin{equation} \label{eqn:rkhs}
(\widehat{\mathbf{f}}^1,\cdots,\widehat{\mathbf{f}}^D)=    \min_{\mathbf{f}^1 \in \mathcal{H}({\mathcal{\mathbf{K}}}^1),\ldots,\mathbf{f}^d \in \mathcal{H}(\mathcal{\mathbf{K}}^D)}\sum_{d=1}^{D}\sum_{j=1}^{r}\frac{1}{2n}\sum_{i=1}^{n}({g}_{i,j} - f_j^d(\bx^d_i))^2 + \frac{\lambda^d}{2} \sum_{d=1}^{D}\|\mathbf{f}^d\|^2_{\mathcal{H}(\mathcal{\mathbf{K}}^d)}.
\end{equation}
Here,  $g_{i,j}$ is the $ij$th element of $\bG$, $\bx_i^d$ is the $i$th row of $\bX^{n \times p^d}$ and  $\|\mathbf{f}^d\|_{\mathcal{H}(\mathbf{K^d})}$ is the norm induced by the inner product in a RKHS. The first term in equation (\ref{eqn:rkhs}) is the empirical loss when predicting $f_j^d(\bx^d_i)$ in place of $g_{i,j}$, and $\lambda^d >0$ is a regularization parameter that controls the trade-off between model fit and  complexity. 
Minimizing  (\ref{eqn:rkhs}) is equivalent to minimizing $D$ independent equations:
\begin{equation} \label{eqn:rkhs2}
\widehat{\mathbf{f}}^d=    \min_{\mathbf{f}^d \in \mathcal{H}(\mathcal{\mathbf{K}}^d)}\sum_{j=1}^{r}\frac{1}{2n}\sum_{i=1}^{n}(g_{i,j} - f_j^d(\bx^d_i))^2 + \frac{\lambda^d}{2} \|\mathbf{f}^d\|^2_{\mathcal{H}(\mathcal{\mathbf{K}}^d)}, d=1,\ldots,D.
\end{equation}
From the representer theorem of RKHS \citep{alvarez2011kernels, kimeldorf1970correspondence}, the solution $\widehat{\mathbf{f}}^d$ at any point $\widetilde{\bx}^d \in \Re^{p^d}$ can be represented as a finite linear combination of the kernel matrix $\mathcal{\mathbf{K}}^d$ over the $n$ data points:
\begin{equation} \label{eqn:rkhssol}
\widehat{\mathbf{f}}^d(\widetilde{\bx}^d)= \sum_{i=1}^{n}\mathcal{\mathbf{K}}^d(\bx_i^d, \widetilde{\bx}^d)\balpha_i^d,~~ \balpha_i^d \in \Re^{r},
\end{equation}
where we note that $\mathcal{\mathbf{K}}^d(\bx_i^d, \widetilde{\bx}^d)$ is an $r\times r$ matrix acting on the $r$-long vector $\balpha_i^d$. 
The parameter $\balpha^d$, $d=1,\ldots,D$ for all subject combined satisfies the linear system
\begin{equation} \label{eqn:rkhssol2}
\widetilde{\balpha}^d = (\mathcal{\mathbf{K}}^d(\bX^d,\bX^d) + \lambda^d\bI_{nr})^{-1}\widetilde{\bg},
\end{equation}
where $\widetilde{\balpha}^d$ is a  length-$nr$ vector.  Of note, $\widetilde{\bg}$ is also a length-$nr$ vector obtained by concatenating the columns in $\bG$; $\mathcal{\mathbf{K}}^d(\bX^d,\bX^d)$ is an $nr \times nr$ block matrix consisting of blocks of size  $n \times n $.

Note that to obtain $\widetilde{\balpha}^d$, we must invert the $nr \times nr$ matrix $\mathcal{\mathbf{K}}^d(\bX^d,\bX^d)$ , which can be computationally expensive for larger samples. We proceed as follows for computational ease. We assume that the $r$ components in the $\bG$ output matrix are independent and therefore uncorrelated. This assumption is reasonable and enables us to obtain unique information about each component in $\bG$. We also assume the kernel $\mathbf{K}^d$ is \textit{separable}.  That is $\mathbf{K}^d$ can be written as a Kronecker product of two kernels $\mathcal{K}^d\otimes \mathcal{B}$. Here, $\mathcal{B}$ is an $r \times r$ symmetric and positive semi-definite matrix capturing dependencies among components in $\bG$, and $\mathcal{K}^d$ is an $n\times n$ kernel matrix for the input space. The kernel matrix $\mathbf{K}^d$ is therefore the product of the function for the input space alone (i.e. $\mathcal{K}^d$) and the interactions between variables in the output space (i.e. $\mathcal{B}$), in our case, between the $r$ components in $\bG$. By assuming that the components are independent, we can  write $\mathcal{\mathbf{K}}^d(\bx_i^d, \widetilde{\bx}^d)$ in equation (\ref{eqn:rkhssol}) as $\mathcal{K}^d(\bx_i^d, \widetilde{\bx}^d)  \mathbf{I}$ \citep{alvarez2011kernels}, where $\mathcal{K}^d(\bx_i^d, \widetilde{\bx}^d)$ is a scalar kernel, and $\mathcal{B}=\mathbf{I}$, an $r \times r$ identity matrix. Thus, the kernel matrix $\mathcal{\mathbf{K}}^d(\bX^d,\bX^d)$ becomes block diagonal \citep{alvarez2011kernels}. Since we use the same input data to estimate the $j$th and $j'$th component, all blocks are equal. Therefore, the inversion of the $nr \times nr$ matrix is reduced to inverting the $n \times n$ matrix $\mathcal{K}^d (\bX,\bX)$.


With these concepts, the functional in equation (\ref{eqn:rkhs2}) can be written as $\sum_{j=1}^r\|f^d_j \|^2_{\mathcal{H}(\mathcal{K}^d)}$, and the minimization problem (\ref{eqn:rkhs2}) reduces to the minimization of $r$ independent functionals:
\begin{equation} \label{eqn:rkhs2r}
(\widehat{{f}}^d_1,\ldots,\widehat{{f}}^d_r)=    \min_{f^d_1 \in \mathcal{H}(\mathcal{K}^d),\cdots,f^d_r \in \mathcal{H}(\mathcal{K}^d)}\sum_{j=1}^{r}\frac{1}{2n}\sum_{i=1}^{n}(g_{i,j} - f_j^d(\bx^d_i))^2 + \frac{\lambda^d}{2} \sum_{j=1}^r\|f^d_j \|^2_{\mathcal{H}(\mathcal{K}^d)}, d=1,\ldots,D.
\end{equation}
Correspondingly, equations (\ref{eqn:rkhssol}) and (\ref{eqn:rkhssol2}) for problem (\ref{eqn:rkhs2r}) become 
\begin{equation} \label{eqn:rkhssolr}
\widehat{{f_j}}^d(\widetilde{\bx}^d)= \sum_{i=1}^{n}\mathcal{K}^d(\bx_i^d, \widetilde{\bx}^d)\alpha_{i_j}^d,~~
\widetilde{\balpha}^d_j = (\mathcal{{K}}^d(\bX^d,\bX^d) + \lambda^d\bI_{n})^{-1}\widetilde{\bg}_j ~~~~\mbox{for~$j=1,\ldots,r$,~}
\end{equation}
where $\widetilde{\balpha}^d_j$ and $\widetilde{\bg}_j$ are both vectors of length $n$, and $\widetilde{\bg}_j$ is the $j$th column in $\bG$. Here, $\mathcal{K}^d(\mathbf{X}^d,\mathbf{X}^d)$ is  an $n \times n$ matrix with the $ij$th entry $\mathcal{K}^d(\mathbf{x}_i^d, \mathbf{x}^d_j)$. 

\subsubsection{Random Fourier Transforms}
 Inverting the $n \times n$ matrix $\mathcal{{K}}^d(\bX^d, \bX^d)$ can still be expensive for large $n$. To reduce computations further, we follow ideas in \cite{rahimi2008random} and use random Fourier transforms to construct low-dimensional approximations of our data. 
Specifically, we map the data into a low-dimensional euclidean inner product  space via a randomized feature map, $\bz: \Re^{p^d} \rightarrow  \Re^M$ such that the inner product of two transformed points  in this space closely approximate the corresponding kernel evaluation: 
$ \mathcal{K}^d(\bx^d,\widetilde{\bx}^d) \approx \langle\ \bz(\bx^d),\bz(\widetilde{\bx}^d) \rangle_{\Re^M}=\bz(\bx^d)^{\smt}\bz(\widetilde{\bx}^d)$ \citep{rahimi2008random}.
Using the nonlinear features $\bz(\bx^d)$, the solution to the minimization problem in equation (\ref{eqn:rkhs2r}), which is given by equation (\ref{eqn:rkhssolr}), can be approximated by:
\begin{equation} \label{eqn:rf}
\widehat{f}_j^d(\widetilde{\bx}^d) \approx \langle\ \bz(\bx^d),\bbeta_j^d\rangle_{\Re^M}\\,~~
(\mathbf{Z}^{d^\smt}\mathbf{Z}^{d} + \lambda^d\bI_{M})\bbeta_j^d = \mathbf{Z}^{d^\smt}\bg_j, \forall j,
\end{equation}

where $\mathbf{Z}^{d}$ is an $n \times M$ matrix, and  $\bbeta_j^d$ is a length-$M$ vector. Let $\bA^d=(\bbeta_1^d,\ldots,\bbeta_r^d) \in \Re^{M \times r}$ be the collection of all $r$ coefficients $\bbeta_j^d, j=1,\ldots,r$, for view $d$. Then, $\mathbf{A}=(\mathbf{Z}^{d^\smt}\mathbf{Z}^{d} + \lambda^d\bI_{M})\mathbf{Z}^{d^\smt}\mathbf{G}$. The solution  requires us to invert an $M \times M$ matrix instead of $n \times n$, which is computationally efficient since $M \ll n$. 

To construct a well-approximating feature map, the authors \cite{rahimi2008random} invoked Bochner's theorem \citep{rudin2017fourier}, which states that each entry of a shift-invariant kernel matrix $\mathcal{K}(\mathbf{X},\mathbf{X}) \in \Re^{n \times n}$ (we suppress the superscript $d$)  can be approximated as:
$\mathcal{K}(\bx_l-{\bx_j}) =\int p(\mathbf{w})e^{i\mathbf{w}^{\smt}(\mathbf{x}_l- \mathbf{x}_j)}d(\mathbf{w})  
 \approx \frac{1}{M}\sum_{m=1}^{M}\langle\sqrt{2}\cos(\bw ^{\smt}_m\bx_l + b_m),\sqrt{2}\cos(\bw ^{\smt}_m\bx_j + b_m) \rangle 
 = \langle~\mathbf{z}(\mathbf{x}_l), \mathbf{z}(\mathbf{x}_j)~ \rangle. $
Here, $\mathbf{z}(\mathbf{x}_l)=[\sqrt{2}\cos(\bw^{\smt}_1\bx_l + b_1),\cdots,\sqrt{2}\cos(\bw ^{\smt}_M\bx_l + b_M)] \in \Re^{M}$, 
$p(\mathbf{w})$ is the inverse Fourier transform of the shift-invariant kernel $\mathcal{K}(\cdot)$, $d(\mathbf{w})$ is a probability measure on $\bw$; ${w}_1,\ldots,{w}_M$ are sampled from $p(\mathbf{w})$,  and $b_m \sim Unif(0, 2\pi)$. 
Essentially, to obtain each component of $\bz(\bx^d)$, the data $\bx^d$ are projected onto a random direction $\bw^d$ drawn from $p(\mathbf{w})$, then they are rotated by a random amount $b^d$, and the resulting scalar is passed through a sinusoidal function.  For the Gaussian kernel $\mathcal{K}(\mathbf{x}_l,\mathbf{x}_j)= \exp(\frac{-\|\bx_l - \bx_j\|_2^2}{2\nu^2})$, the inverse Fourier transform, $p(\mathbf{w})$, is given by a Gaussian distribution, i.e.  $\mathbf{w}_{m}^d \sim N(\mathbf{0},\frac{1}{\nu^2}\bI)$ \citep{yang2012nystrom}. 

\subsubsection{Variable selection and optimization problem for nonlinear association of multiview data}\label{sec:varsel}
One of our main objectives is to identify important variables in $\bX^d$, which underlie the low-dimensional structure of the views as well as affect the change in the outcome. The task here is to ensure that all $M$ random samples of $\bw^d$, $d=1,\ldots,D$, have the same sparse structure. We use the reparametrization approach proposed in  \cite{gregorova2018large} to re-write $\bw^d$ as $\bw^d=\mathbf{\epsilon}^d\cdot \bgamma^d$, $\mathbf{\epsilon}^d$  is drawn from $p(\mathbf{w})$, and $\cdot$ is the element-wise product.  Thus, we use $\bgamma^d$ to scale each feature, and  consider random feature mapping:
\[\bz(\bx^d)=[\sqrt{2}\cos(\mathbf{\epsilon}_1^{d\smt}(\bgamma^d\cdot\bx^d) + b_1^d), \sqrt{2}\cos(\mathbf{\epsilon}_2^{d\smt}(\bgamma^d\cdot\bx^d) + b_2^d), \ldots, \sqrt{2}\cos(\mathbf{\epsilon}_M^{d\smt}(\bgamma^d\cdot\bx^d) + b_M^d)]. \]
Therefore, sparsity in $\mathbf{w}^d$ (and thus the learned model) for view $d$ is achieved by sparsity in $\bgamma^d$. 
Like feature scaling methods, we propose a multi-step alternating procedure to learn model parameters $\mathbf{A}$ and $\gamma^d$, for a fixed $\mathbf{G}$. For a fixed $\gamma^d$ and a fixed $\mathbf{G}$, we generate random features for all the input data points for each view and solve the linear problem \eqref{eqn:rf} to get the $M \times r$ matrix $\mathbf{A}$. With this fixed, we solve the $p^d$-length vector $\gamma^d$ of feature scalings. Since we do not know $\mathbf{G}$, we estimate it.  
We formulate the following optimization problem to learn the common low-dimensional nonlinear representation $\bG$, the matrix of coefficients $\bA^d$ and the sparse weights $\gamma^d$:
\begin{eqnarray} \label{eqn:assoc}
    \min_{\bG,\bgamma^1,\ldots \bgamma^D,\bA^1,\cdots, \bA^D} \left\{\frac{1}{2n}\sum_{d=1}^{D}\|\bG - \bZ^d\bA^d\|^2_F +\frac{\lambda^d}{2} \sum_{d=1}^{D}\|\bA^d\|^2_F  + \sum_{d=1}^{D}\mathcal{P}(\bgamma^d)\right\} \nonumber\\
    \mbox{subject~ to}~  \bG^{\smt}\bG=\bI_r,  
\end{eqnarray}
where it is clear that $\mathbf{Z}^d$ depends on $\bgamma^d$. We consider two options for sparsity in each view: individual variable sparsity and group sparsity. For individual variable sparsity (i.e. we allow coefficients of individual variables to be zero), we restrict $\bgamma^d$ to lie in a probability simplex ($\Delta$)  i.e. $\bgamma^d \in \Delta$. $\bgamma^d$ lies in a probability simplex if $0 \le \gamma^d_s \le 1, s=1,\ldots,p^d$, and $\sum_{s=1}^{p^d}\gamma^d_s =1$. 
For group sparsity, we consider the non-overlapping sparse group penalty: $\mathcal{P}(\bgamma^d) = \eta\rho \|\bgamma^d\|_{1} + (1-\eta)\rho \sum_{l=1}^{G^d}\|\sqrt{p_l^d}\gamma_l^d\|_{2}$, where $\|\cdot\|_1$ is the $l_1$ norm, $\| \cdot\|_2$ is the $l_2$ norm, $G^d$ is the number of non-overlapping groups for view $d$ that is known a prior, $p_l^d$ is the number of variables in group $l, l=1,\ldots,G^d$,  $\eta \in [0,1]$ combines the lasso and group lasso penalties, and $\rho >0$ is a sparsity parameter controlling the amount of sparsity for view $d$, for a fixed $\eta$. This penalty allows to select groups as well as individual variables within groups that drive the underlying common low-dimensional structure in the views and  the variation in the outcome.  For a fixed $\rho$, the parameter  $\eta$ balances variable selection of a group overall (i.e. number of groups with at least one nonzero coefficient) and  within group (i.e. number of nonzero coefficients within a nonzero group). Smaller  values of $\eta$ encourage grouping (i.e. more nonzero groups are selected) and individual variable selection within groups (i.e. more variables tend to have nonzero coefficients within groups)  while larger variables of $\eta$ discourages group selection and encourages sparsity within groups (i.e. more zero coefficients within groups). 

\subsection{Supervised nonlinear association studies}
In this section, we place our research in the context of methods that simultaneously associate data from multiple views and predict an outcome. 
One of our main goals is to leverage an outcome to guide the construction of a common low-dimensional nonlinear representation and detection of variables that affect the underlying structure in the views.  For this purpose, we assume that the outcome $\bf{y}$, or $\mathbf{Y}$ for multiple continuous outcomes (or multi-class problem), depends on the $d$th view $\bX^{d}$ through the common factor $\bG$. This allows integrative analysis to be combined with the prediction of a clinical outcome simultaneously.  In particular, $\bG$ is associated with an outcome by minimizing the loss function: 
$\mathcal{L}(\mathbf{Y}, \bG, \mathbf{\Theta}~ \mbox{or}~ \btheta)$, 
where $\btheta$, or $\mathbf{\Theta}$ is the effect of the shared factors on $\mathbf{y}$ or $\mathbf{Y}$, respectively. We formulate the following optimization problem to model nonlinear association of  multiple views, predict an outcome or outcomes and select relevant individual or groups of variables: 
\begin{eqnarray}\label{eqn:overallopt}
 \min_{\bG,\mathbf{\Theta},\ldots, \bgamma^1,\ldots \bgamma^D,\bA^1,\cdots, \bA^D} \left\{\mathcal{L}(\mathbf{Y}, \bG, \mathbf{\Theta}) +  \frac{1}{2n}\sum_{d=1}^{D}\|\bG - \bZ^d\bA^d\|^2_F +\frac{\lambda^d}{2} \sum_{d=1}^{D}\|\bA^d\|^2_F + \sum_{d=1}^{D}\mathcal{P}(\bgamma^d)\right\} \nonumber\\
    \mbox{s.t}~  \bG^{\smt}\bG=\bI_r.
\end{eqnarray}
\noindent For a continuous outcome, $\mathcal{L}(\mathbf{y}, \bG, \btheta)=\frac{1}{2n}\|\mathbf{y} - \bG\btheta\|^{2}_{F}$ and $\| \cdot\|_{F}$ is the Frobenius norm. We assume the outcome data is centered so that there is no need for an intercept term. For multiple continuous outcomes, we let $\mathcal{L}(\mathbf{Y}, \bG, \mathbf{\Theta})=\frac{1}{2n}\|\mathbf{Y} - \bG\mathbf{\Theta}\|^{2}_{F}$, with $\mathbf{\Theta} \in \Re^{r \times q}$.    

For a categorical outcome with number of classes $K\ge 2$, we consider a linear discriminant analysis (LDA) problem. In particular, we use the optimal scoring approach proposed in \cite{opscore:1994} that yields the LDA classification rule. The optimal scoring approach transforms a  classification problem into a regression problem by using a sequence of scores to  convert categorical variables into continuous variables. Let $\mathbf{W}$ be an $n \times K $ indicator matrix with $1$ in the $i$th row and $K$th column if sample $i$ belongs to class $K$, and $0$ otherwise. Let $\widetilde{\bB}$  be a $K \times {K-1}$ matrix of \textit{optimal scores}, and 
$\bar{\mathbf{Y}}= \mathbf{W}\mathbf{B} \in \Re^{n \times (K-1)}$ be the transformed multi-class response. The optimal scoring approach minimizes the objective function  $\mathcal{L}(\bar{\mathbf{Y}}, \bG, \mathbf{\Theta})=\frac{1}{2n}\|\bar{\mathbf{Y}} - \bG\mathbf{\Theta}\|^{2}_{F}$ subject to $\bar{\mathbf{Y}}^{\smt}\bar{\mathbf{Y}}=\mathbf{B}^{\smt}\mathbf{W}^{\smt}\mathbf{W}\mathbf{B}=n\mathbf{I}_{K-1}$ and $\bar{\mathbf{Y}}^{\smt}\mathbf{W}\mathbf{1} = \mathbf{B}^{\smt}\mathbf{W}^{\smt}\mathbf{W}\mathbf{1}=\mathbf{0}$, where $\mathbf{1} \in \Re^{K}$ is a vector of ones, and  $\mathbf{0} \in \Re^{K}$ is a vector of zeros. 
Here, $\mathbf{\Theta}$ is an $r \times K-1$ coefficient matrix. Without the need for any ambiguity, we will clearly define whether $\mathbf{\Theta}$ is the coefficient matrix  of the categorical outcome  or the multiple continuous outcomes. 
Let  $\mathbf{B}_l \in \Re^{K}, l=1,\ldots,K-1$ be  the $l$th column in $\mathbf{B}$, $n_k$ be the number of samples for class $k$ and $s_k=\sum_{i=1}^{k}n_{i}$. Then $\mathbf{B}_l$ can be defined  in terms of the sample size $n_k$ \citep{gaynanova2020prediction} as:
$\mathbf{B}_l = \left( \left\{(nn_{l+1})^{1/2}(s_ls_{l+1})^{-1/2}\right\}_l,~~ -(ns_l)^{1/2}(n_{l+1}s_{l+1})^{-1/2},~~ \mathbf{0}_{K-1-l}\right)^{\smt}$. 

\subsection{Predicting an  outcome}
In this Section, we describe our approach to predict an outcome for  a given target data, $\mathbf{X}_{target}^d$. 
Using the training data, we solve the optimization problem in equation  \ref{eqn:overallopt} and we use the estimates  $\widehat{\bgamma}^d$, $\widehat{\bA}^d, d=1,\ldots,D$, and $\widehat{\mathbf{\Theta}}$ (or $\widehat{\mathbf{\btheta}}$) and the target data to predict the outcome. We first construct the target nonlinear features for subject $i$ as \[\bz(\bx^d_{target})_{target}=[\sqrt{2}\cos(\mathbf{\epsilon}_1^{d\smt}(\widehat{\bgamma}^d\cdot\bx^d_{target}) + b_1^d),  \ldots, \sqrt{2}\cos(\mathbf{\epsilon}_M^{d\smt}(\widehat{\bgamma}^d\cdot\bx^d) + b_M^d)], \]
where $\widehat{\bgamma}^d$ is the learned sparse pattern for view $d$ and $\bx^d_{target}$ is a length-$p^d$ target data for subject $i$. We note that we use the same $\epsilon_j^d$ and $b^d_j, j=1,\ldots,M$ samples that were constructed for the training data. Let $\mathbf{Z}_{target} \in \Re^{n_{target} \times M}$ be a matrix  with rows $\bz(\bx^d_{target})$. Given $\mathbf{Z}_{target}$, we solve the following optimization problem to obtain the target shared low-dimensional representation:
 $\widehat{\bG}_{target}=   \min_{\bG: \bG^{\smt}\bG = \bI_r} \frac{1}{2n}\sum_{d=1}^{D}\|\bG - \bZ^d_{target}\widehat{\bA}^d\|^2_F$. 
The solution to this problem can easily be obtained. Now, given the estimate $\widehat{\mathbf{G}}_{target}$, we predict a continuous outcome $\mathbf{y}_{pred}$ as $\mathbf{y}_{pred} = \widehat{\bG}_{target}\widehat{\btheta}$ for a single continuous outcome, or $\bY_{pred}=\widehat{\bG}_{target}\widehat{\mathbf{\Theta}}$ for multiple continuous outcomes. For a categorical outcome,  we obtain the target scores $\mathbf{\bU}_{target} = \widehat{\mathbf{G}}_{target}\widehat{\mathbf{\Theta}}$ and the training scores $\mathbf{\bU} = \widehat{\mathbf{G}}\widehat{\mathbf{\Theta}}$. We then use the nearest centroid algorithm to assign each subject to the class that minimizes the difference between the target score for that subject and class average from training scores. 

\section{Algorithm}\label{sec:Alg}

The optimization of (\ref{eqn:overallopt}) is done by alternating minimization over $\mathbf{A}^d$, $\mathbf{G}$, $\mathbf{\Theta}$ or $\btheta$, and $\bgamma^d$. Refer to the Supporting Information for more details on our initializations, Algorithms, and hyperparameters selection.
We  estimate each parameter  as follows.
 
\textit{Estimate view-specific loadings, $\mathbf{A}^d$}: 
To estimate $\mathbf{A}^d$ at iteration $t+1$, we fix $\bG$ and $\bgamma^d$ at their values at iteration $t$, $t\ge 0$, and we solve the minimization problem:\\
  $  \min_{\bA^1,\cdots, \bA^D} \left\{\frac{1}{2n}\sum_{d=1}^{D}\|\bG - \bZ^d\bA^d\|^2_F + \frac{\lambda^d}{2} \sum_{d=1}^{D}\|\bA^d\|^2_F \right\}$.
Let $\widetilde{\bZ}^d = \left( \frac{1}{\sqrt{n}}\bZ^d ,~~ \sqrt{\lambda^d}\bI_{D \times D}\right)_{(n+D) \times D}$. Also let $\widetilde{\bG} = \left(\frac{1}{\sqrt{n}}\bG~~~ ,\mathbf{0}_{D \times r} \right)_{(n+D) \times r}$. The minimization problem may be written as:\\
  $  \min_{\bA^1,\cdots, \bA^D} \left\{\frac{1}{2}\sum_{d=1}^{D}\|\widetilde{\bG} - \widetilde{\bZ}^d\bA^d\|^2_F. \right\}$.
This reduces to minimizing $D$ independent least squares optimization problems for $\bA^d, d=1,\ldots,D$, and each has a closed form solution given by $\widehat{\bA}^d= (\widetilde{\bZ}^{d^\smt}\widetilde{\bZ})^{-1}\widetilde{\bZ}^{d^\smt}\widetilde{\bG}$.

\textit{Estimate shared low-dimensional representation, $\mathbf{G}$}: 
For $\bA^d$ fixed at its value from iteration $t+1$, and for $\mathbf{\Theta}$ (or $\btheta$) fixed at their values in iteration $t$, we solve for $\widehat{\bG}$ at iteration $t+1$:
$    \min_{\bG:\bG^{\smt}\bG=\bI_r} \left\{\mathcal{L}(\mathbf{Y}, \bG, \mathbf{\Theta}) +  \frac{1}{2n}\sum_{d=1}^{D}\|\bG - \bZ^d\bA^d\|^2_F  \right\}$.
The solution for $\bG$ can be written in a  closed form. Specifically,  let $\widetilde{\mathbf{Y}} = \left(\mathbf{Y} ~\text{or}~ \bar{\mathbf{Y}}, \bZ^1\bA^{1^{(t+1)}}, \cdots,   \bZ^D\bA^{D^{(t+1)}} \right)_{n \times (Dr +q)}$. For a single continuous outcome, $q=1$. Of note, $\bY$ is the continuous response outcome(s) and $\bar{\bY}$ is the transformed multi-class response. For a $K$-class classification problem, $q=K-1$. Let $\widetilde{\mathbf{\Theta}}=\left(\mathbf{\Theta}, \bI_{r},\cdots,\bI_{r}\right)_{r \times (Dr +q)}$. Then, problem is the same as 
   $ \min_{\bG:\bG^{\smt}\bG=\bI_{r}}\frac{1}{2n}\|\widetilde{\bY} - \bG\widetilde{\mathbf{\Theta}}\|^2_F$,
a Procrustes optimization problem.  But minimizing the above problem is the same as solving the problem:
$\max_{\bG:\bG^{\smt}\bG=\bI_{r}} tr\left(\frac{1}{n}\widetilde{\bY}\widetilde{\mathbf{\Theta}}^{\smt}\bG\right)$,
where $tr(\cdot)$ is the trace function. The solution to the above is  $\bU\bV^{\smt}$ where $\bU$ and $\bV$ are the left and right singular vectors from the spectral decomposition of $\frac{1}{n}\widetilde{\bY}\widetilde{\mathbf{\Theta}}^{\smt} = \bU\bD\bV^{\smt}$.

{\textit{Estimate $\mathbf{\Theta}$  or $\btheta$}}: 
For $\bG$ fixed at its value at iteration $t+1$, we solve for $\widehat{\mathbf{\Theta}}$ (for multiple response or multi-class problems) or $\widehat{\btheta}$ for a single outcome using the optimization problem:  $ \min_{\mathbf{\Theta}~ \text{or}~ \btheta} \mathcal{L}(\mathbf{Y} [\mbox{or}~ \mathbf{y}~ \mbox{or}~ \bar{\bY}], \bG, \mathbf{\Theta})$.
This is a least squares problem and the solution is $\widehat{\mathbf{\Theta}}=(\bG^{\smt}\bG)^{-1}\bY$ for a continuous outcome(s) or $\widehat{\mathbf{\Theta}}=(\bG^{\smt}\bG)^{-1}\bar{\bY}$ for a categorical outcome. 

\textit{Estimate $\bgamma^d$}: 
Recall that $\bZ^d$ depends on $\bgamma^d$. For $\bG$ and $\bA^d$ fixed at their current values, we solve the following problem for view $d$:
 $   \min_{\bgamma^d} \frac{1}{2n}\|\bG - \bZ^d\bA^d\|^2_F  + \mathcal{P}(\gamma^d).$
As mentioned in Section \ref{sec:varsel}, we consider two variable selection options for $\mathcal{P}(\gamma^d)$: (i) individual variable selection where we restrict $\bgamma^d$ to lie in a probability simplex ($\Delta$)  i.e. $\bgamma^d \in \Delta$, and (ii) group variable selection where we use the non-overlapping sparse group penalty: $\mathcal{P}(\bgamma^d) = \eta\rho \|\bgamma^d\|_{1} + (1-\eta)\rho \sum_{l=1}^{G^d}\|\sqrt{p_l^d}\gamma_l^d\|_{2}$. For either option, we solve the optimization problem using accelerated gradient descent. In particular, we implement the fast iterative shrinkage-thresholding  algorithm (FISTA) with backtracking proposed in \citep{beck2009fast}.

\section{Simulations}
\label{s:sim}

We conduct simulation studies to assess the empirical performance of the proposed methods. Our main goal is to assess the accuracy of our method in predicting a test outcome and selecting variables based on the low-dimensional representations  and the sparsity parameter learned. 
We consider two scenarios based on the type of outcome. Each scenario has $D=2$ views and we simulate data with nonlinear relationships between and within views. In the first scenario, low-dimensional representations are used to generate an outcome with continuous values (Supporting Information).  In the second scenario, the data are generated to have two classes. In all scenarios, we generate 20 Monte Carlo training and testing sets. We evaluate the proposed and existing methods using the following criteria: i) test prediction/classification accuracy and ii) feature selection. For prediction, we evaluate the ability of the proposed methods to correctly predict (using mean squared error [MSE] for continuous outcomes) or classify (using classification accuracy for binary outcomes) based on the shared low-dimensional representations learned.   In feature selection, we evaluate the methods ability to choose true signals rather than noise variables.

\subsection{Binary Outcome }
\begin{figure}[htb!]
\begin{center}
\begin{tabular}{cc}
         \includegraphics[width=0.45\textwidth]{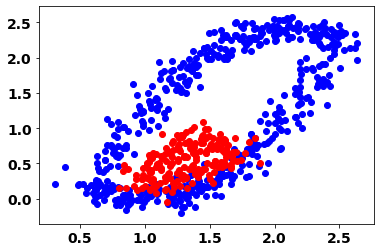}&\includegraphics[width=0.45\textwidth]{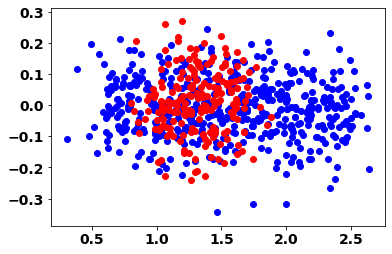}\\
         \includegraphics[width=0.45\textwidth]{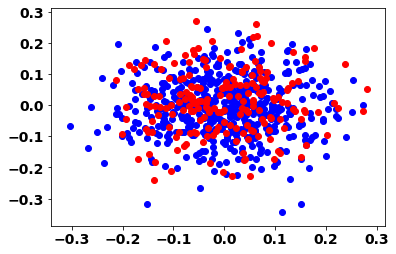}&\includegraphics[width=0.45\textwidth]{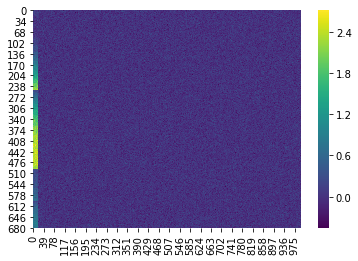}\\
\end{tabular}
\end{center}
\vspace{0.2in}
    \caption{Top Left: Relationship between signal variables 1 and 2 in View 1; Top Right: Relationship between signal variables 1 and noise variable 21 in View 1; Bottom Left : Relationship between noise variables; Bottom Right: Image plot of View 1. First twenty variables are signals and remaining variables are noise. Plots of View 2 are similar to View 1.}
    \label{fig:nonlinearbin}
\end{figure}

We consider three different settings, each with a different number of samples and variables. Twenty variables have nonlinear relationships (Figure \ref{fig:nonlinearbin}) in each setting and are considered signal variables. We generate views with two classes in each view. 
\begin{itemize}
   \item Generate data for class one as follows:
   \begin{itemize}
    \item Generate $\btheta_1$ as a vector of $n_1/2$ evenly spaced points between $0.6$ and $2.5$
    \item Form the vector $\mathbf{s}=[(\btheta_1 -\mathbf{1})^2 ;~ (\btheta_1 +0.1\mathbf{1})^2 -2 (\btheta_1 -\mathbf{1})^2] \in \Re^{n_1 \times 1}$
    \item Set $\widetilde{\bX}_{11} =[\tilde{\btheta}_1; \mathbf{s}\mathbf{1}^{\smt}_{p_1-1} ] \in \Re^{n_1 \times p^1}, \tilde{\btheta}_1=[\btheta_1;\btheta_1] \in \Re^{n_1 \times 1}$ 
    \item Generate $\mathbf{X}_{11}= \widetilde{\bX}_{11} \cdot \bW + \sigma_{11}\bf{E}_{11}$ where $(\cdot)$ is element-wise multiplication,  $\bW \in \Re^{n_1 \times p^1}= [\mathbf{1}_{20}, \mathbf{0}_{p^1-20}]$ is a matrix of ones and zeros, $\mathbf{1}$ is a matrix of ones, $\mathbf{0}$ is matrix of zeros, $\sigma_{11} =0.1$, and $\mathbf{E}_{11} \sim N(0,1)$. 
    \end{itemize}
    \item Generate data for class two as follows: 
     \begin{itemize}
    \item Generate $\btheta_2$ as a vector of $n_2/2$ evenly spaced points between $0.96$ and $1.67$
    \item Form the vector $\mathbf{s}=[(\btheta_2 -\mathbf{1})^2 + 0.25;~ (\btheta_2 +0.1\mathbf{1})^2 -3.5 (\btheta_2 -\mathbf{1})^2 + 0.25] \in \Re^{n_2 \times 1}$
    \item Set $\widetilde{\bX}_{12} =[\tilde{\btheta_2}; \mathbf{s}\mathbf{1}^{\smt}_{p^1-1} ] \in \Re^{n_2 \times p^1}, \tilde{\btheta}_2=[\btheta_2;\btheta_2] \in \Re^{n_2 \times 1}$ 
    \item Generate $\mathbf{X}_{12}= \widetilde{\bX}_{12} \cdot \bW + \sigma_{12}\mathbf{E}_{12}$ where $(\cdot)$ is element-wise multiplication,  $\bW \in \Re^{n_2 \times p^1}= [\mathbf{1}_{20}, \mathbf{0}_{p^1-20}]$ is a matrix of ones and zeros, $\mathbf{1}$ is a matrix of ones, $\mathbf{0}$ is matrix of zeros, $\sigma_{12} =0.1$, and $\mathbf{E}_{12} \sim N(0,1)$.
    \end{itemize}
    \item Concatenate data from the two classes to form data for View 1, i.e. $\bX^1 =[\bX_{11}; \bX_{12}] \in \Re^{n \times p^1}$. 
    \item Generate the second view as:  ${\bX}^2=5\mathbf{X}^1 + \sigma_{2}\mathbf{E}^2$, where $\mathbf{E}_2 \sim N(0,1)$, and $\sigma_2=0.2$. 
    \item By generating $\mathbf{X}_{1}$ and $\mathbf{X}_{2}$ this way, we assume the first $20$ variables have nonlinear relationships and discriminate the two classes. Figure  \ref{fig:nonlinearbin} is a pictorial representation of the relationship between signal variables (left), signal and noise variables (2nd left plot), noise variables (third left plot), and an image plot of data for view 1. 
\end{itemize}

\subsection{Comparison Methods}
We compare the proposed methods, RandMVLearn and RandMVLearnG (for the selection of group variables), with linear and nonlinear methods for associating data from multiple views: sparse canonical correlation analysis [sCCA] \citep{safo2018sparse}, canonical variate regression (CVR) \citep{CCAReg2016}, sparse integrative discriminant analysis for multiview data (SIDA) \citep{SIDA:2019}, BIP \citep{chekouo2023bayesian}, deep canonical correlation analysis (Deep CCA) \citep{Andrew:2013} and MOMA \citep{MOMA:2022}. Since sCCA and Deep CCA are mainly used for associating  views, we use the scores from these methods in linear or nonlinear regression models to investigate the prediction and classification performance of the low-dimensional representations learned from these methods. To examine the advantages of appropriately integrating data from multiple sources, these integrative analysis methods are compared to naive implementations of  multiple layer perceptron (MLP) -- for continuous outcome -- and support vector machine (SVM) \citep{ben2008support} -- for binary outcome -- on stacked views.  We combine Deep CCA with the Teacher-Student Framework (TS) \citep{TS:2019} for variable selection. 

We compare the performance of the proposed variable selection options- individual and group variable selections.
For group variable selection, there are two groups of variables: a group of signal variables and a group of noise variables.  We fix $\eta=0.5$ in the sparse group lasso penalty. We consider $5$ values of $\rho$ in the range $(0, \rho_{max})$ for each view. $\rho_{max}$ was chosen to prevent trivial solutions in $\gamma^d_{SGLasso}$. We perform a grid and random search of plausible combinations of $\rho$ from each view, and choose the optimal $\rho$ combination via three-fold cross-validation.   


\subsection{Results}
We first compare results for the binary outcome. Figure 2 gives the prediction and variable selection performance of the proposed and comparison methods, combined across views and averaged over 20 repetitions.  The variable selection performance for RandMVLearnG (both grid and random search) is better than that of RandMVLearn, highlighting the advantages of using group information when it is available.  RandMVLearnG had higher true positives and lower false positives compared to CVR and MOMA, and competitive variable selection performance compared to sCCA. We noticed that the proposed methods, especially RandMVLearnG, have lower classification error rates than all other methods, in all settings. For the continuous settings (see Figure 2 in Supporting Information), the MSEs for RandMVLearn are comparable to that of the group variable selection (i.e., RandMVLearnG). Further, the MSEs for our proposed methods are comparable (Settings 1, 2 and 4) or better (Setting 3) than the methods compared. The variable selection performance of RandMVLearnG is comparable to sCCA in all settings and better than  CVR. In terms of MSE, sCCA followed by linear regression resulted in lower MSE's in all settings, except in Setting 3. RandMVLearnG using a Random search  yields comparable estimates with RandMVLearnG that uses a grid search for hyperparameters tuning. However, RandMVLearnG (Random) is faster than RandMVLearnG (Grid). Further, RandMVLearn is computationally efficient compared to the other methods. 

In general, the prediction and variable selection results for binary and continuous outcomes, with or without using prior group information, show that our methods can detect true signals and omit noise variables in the data. The proposed methods provide competitive or better prediction estimates even in situations where the sample size is less than the number of variables. 

\begin{figure}[htb!]\label{fig:binarym}
\centering
\includegraphics[scale=0.65]{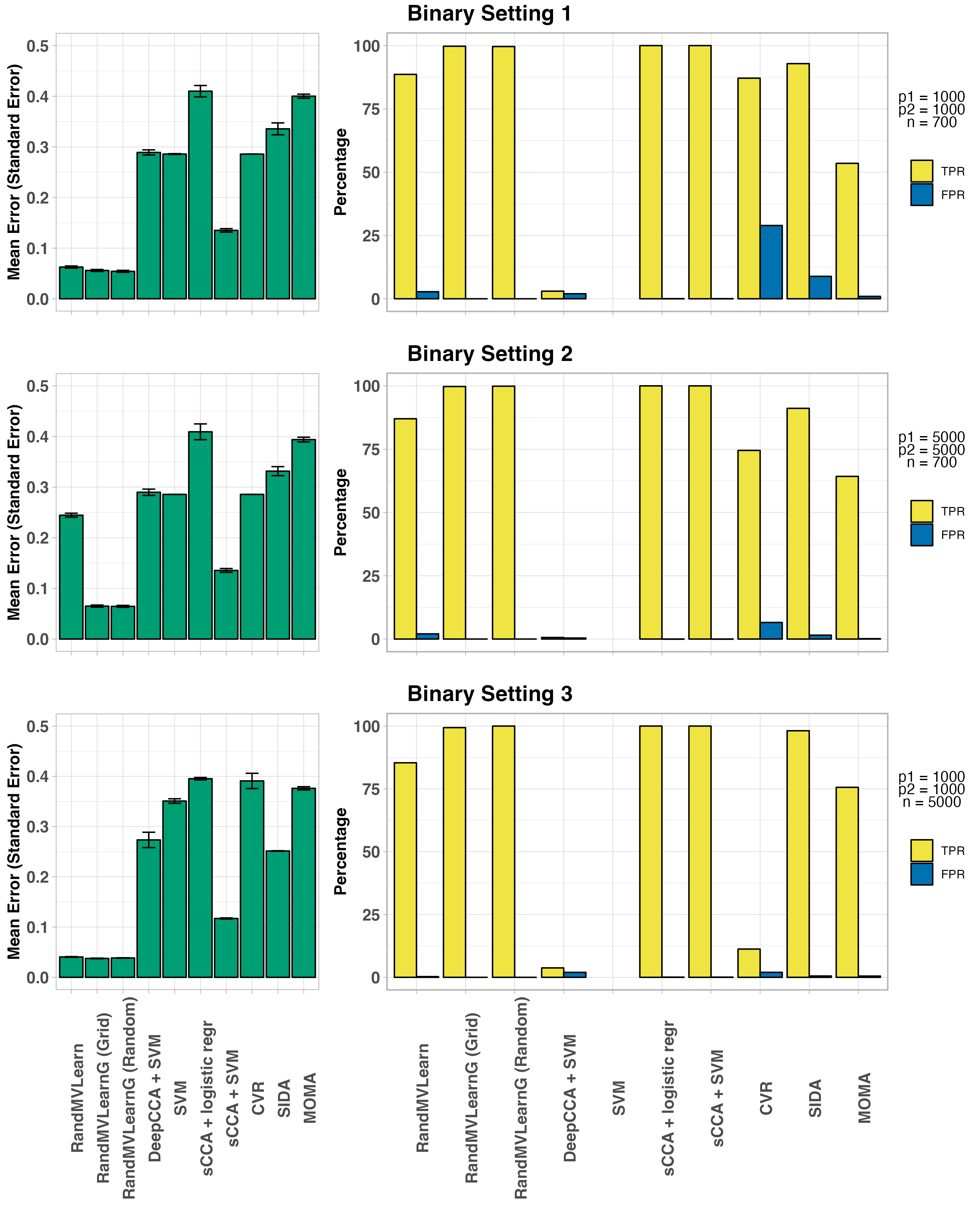}
\caption{Simulation results for binary settings. Number of random features, $M$ set to 300. $(n_1,n_2) = (500, 200), (500, 200), (3000, 2000)$ for Settings 1, 2, and 3, respectively.  Number of components fixed at $K=5$. The proposed methods RandMVLearn, RandMVLearnG have lower error rates in almost all settings, higher or comparable true positive rates (TPR), and lower or comparable false positive rates (FPR)}.
\end{figure}

\section{Analysis of Data from the COVID-19 Study }
We applied the proposed method to integrate proteomics, RNA-seq, lipidomics, and metabolomics data from our motivating study. Our goal is to model nonlinear associations in the molecular and outcome data and to identify molecular signatures and pathways that can contribute to the severity (defined by HFD-45) and status of COVID-19. 

\subsection{Data pre-processing and application of proposed and competing methods}
Of the 128 patients, 120 had both omics and clinical data. Our main outcomes were HFD-45 (continuous outcome) and COVID-19 status (binary outcome). The initial datasets contained $18,212$ genes, $517$ proteins, and $111$ metabolomics and $3,357$ lipidomics features. We used the pre-processed data from \cite{Danietal:2022}. Refer to \cite{Danietal:2022} for more details. We proceeded in two ways to investigate the proposed methods. First, we applied our method without group information on the pre-processed data: $\mathbf{X}^{1} \in \Re^{120 \times 1,015}$ for lipidomics, $\mathbf{X}^{2} \in \Re^{120 \times 72}$ for metabolomics, $\mathbf{X}^{3} \in \Re^{120 \times 5,800}$ for RNA-Seq and $\mathbf{X}^{4} \in \Re^{120 \times 264}$ for proteomics. In the Supporting Information, we consider the sparse  group implementation of the proposed method assuming that group information exists for the RNA-Seq data. 
For both implementations, data were randomly divided into 50 training sets ($n=71$) and testing sets ($n=49$)  while maintaining proportions of COVID-19 status similar to those of the complete data. We used the training data to fit the models.  We used the testing data to assess error rates--MSE and misclassification rates for continuous and binary outcomes, respectively.  We set the number of random features to 45, to be smaller than the training and testing sizes. We chose the number of shared  low-dimensional representations using the simple approach proposed. We chose the kernel parameter using median heuristic. 


\subsubsection{Prediction estimates and molecules selected when there's no group information}
\begin{figure}[htb!]
\begin{center}
\begin{tabular}{cc}
         \includegraphics[scale=0.7]{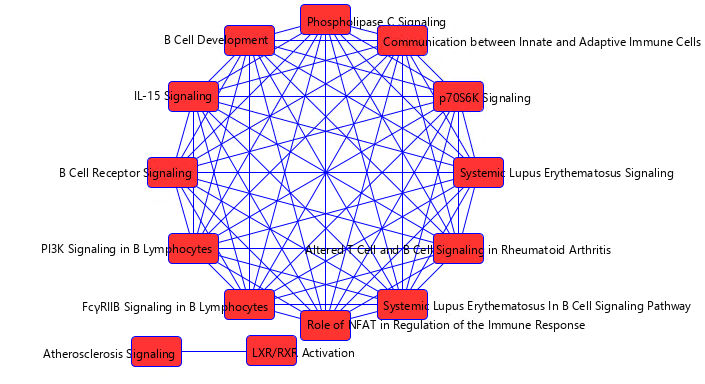}\\
         \includegraphics[scale=0.7]{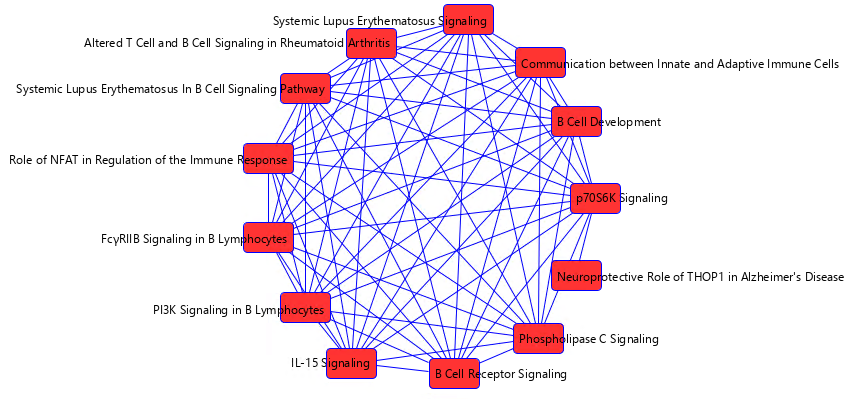}\\
  \end{tabular}
\end{center}
\vspace{0.2in}
    \caption{Overlapping canonical pathways in IPA is used to visualize common molecules involved in the Top 15 significantly (p-value $<$ 0.05) enriched pathways in our protein list. Top: Overlapping Pathways for COVID-19 Status. Bottom: Overlapping Pathways for COVID-19 Severity (HFD-45). While there are many pathways that overlap for COVID-19 severity and status, there are some few unique pathways. }
    \label{fig:ipa}
\end{figure}

We discuss results for when there is no prior group information. Table \ref{tab: covid19} gives the average test misclassification rate for COVID-19 status, and average test MSEs for HFD-45 \\
\textit{\textbf{Binary Outcome: COVID-19 Status}}:
SIDA had the lowest average test error (6.86\%)  rate followed by the proposed method (9.76\%), with Deep GCCA yielding the worst average test error rate (18.08\%). In general, the proposed method selected more variables compared to SIDA. The average number of genes, proteins, metabolomics and lipidomics features selected by the proposed method was 129.98, 170.56, 114.9, and 48.72 respectively. We further explored the variables that were selected more than 25 times ($> 50$\%) out of the 50 replicates. Of these variables, 12 lipidomics features,  55 metabolomics features, and 62 proteins were identified. No gene met this criteria. For insight into the functional classification of the proteins (we focus on proteins due to space restrictions), we performed functional enrichment analysis using Ingenuity Pathway Analysis (IPA) software.   Significantly enriched pathways in our protein list for COVID-19 status are found in Table 2 of the Supporting Information. Many of the enriched pathways are immunology pathways. For instance, B Cell Development, IL-15 Signaling, IL-12 Signaling and Production in Macrophage. These pathways regulate the processes or mechanisms  that contribute to the development of diseases. The molecules IGHV1-18,IGHV1-24,IGHV1-3,IGHV2-70,IGKV1-17, IGKV2-30,IGLV3-19,IGLV4-69 appear to be common for the immunology pathways. 
Pathways unrelated to immune function included the p70S6K Signaling pathway,  LXR/RXR activation pathway, the Atherosclerosis signaling pathway, and the Neuroprotective Role of THOP1 in Alzheimer's Disease. The LXR/RXR activation pathway plays a key role in the regulation of lipid metabolism, inflammation, and cholesterol to bile acid catabolism.  The Atherosclerosis signaling pathway is noteworthy because atherosclerosis  studies have observed similarities and differences between atherosclerosis and COVID-19 \citep{18}. Overlapping canonical pathways (Figure \ref{fig:ipa}) in IPA was used to visualize the shared biology in pathways through the common molecules participating in these pathways. Diseases and biological functions that are over-represented in our protein list include (B-H adjusted pvalue $\le 0.01$): Cardiovascular Disease (e.g. occlusion of  artery, atheroscleosis); Organismal Injury and Abnormalities (e.g. progressive neurological disorder, low grade prostate cancer); Inflammatory Response (e.g. cutaneous lupus erythematosus) and Immunological Disease (e.g. Hodgkin lymphona). There's evidence in the literature to suggest that individuals affected by some of these diseases tend to have severe COVID-19 outcomes.




\textit{\textbf{Continuous Outcome: HFD-45}}:
When the outcome was hospital free days, the proposed method had the lowest MSE compared to the nonlinear method Deep GCCA. BIP had the lowest MSE. The average number of genes, proteins, metabolomics and lipidomics features selected by the proposed method was 215.98, 109.66, 47.8, and 144.98, respectively.  Of these variables,  57 metabolomics features and 54 proteins were selected more than 25 times. No  gene nor lipidomic feature met this criteria. Similar to COVID-19 status, we used IPA to determine significantly enriched pathways and diseases in our protein list. Many of the pathways determined to be related to COVID-19 status were also determined to be related to COVID-19 severity (Table 3 in Supporting Information). Diseases and biological functions that are over-represented in our protein list include (B-H adjusted pvalue $\le 0.01$): Inflammatory Response, Metabolic Disease (e.g. Alzheimer disease, glucose metabolism disorder), Neurological Disease (e.g. Alzheimer disease, progressive neurological disorder) and  Organismal Injury and Abnormalities.  
 The fact that many diseases are enriched in our protein list for COVID-19 severity and status suggest that COVID-19 disrupts many biological systems,  heightening  the need to study the post sequelae effects of this disease  to better understand the mechanisms and to develop
effective treatments.

\begin{table}
\caption{Prediction Estimates and Variable Selection. Deep GCCA + MLP refers to Deep Generalized CCA followed by  Multilayer Perceptron (MLP); MLP is Multilayer Perceptron applied on stacked views;   L: Lipidomics; M:Metabolomics; R: RNA-Seq ; P: Proteomics; MSE refers to mean square error}
\label{tab: covid19}
\begin{center}
\begin{small}
\begin{tabular}{lrr}
\hline
Method& Prediction (Std Error) & \# of variables selected	\\
~ & ~ & (L,M,R,P)\\
			\hline
			\hline
Continuous Outcome (HFD-45) &Average MSE & ~\\
\hline
    RandMVLearn  & 0.8872 (0.02) & 144.98/1015, 47.8/72, 215.98/5800, 109.66/264\\ 
    Deep GCCA + MLP & 0.9446 (0.22) & - \\ 
    BIP & 0.8376 (0.01) & 320.92/1015, 14.94/72, 4122.54/5800, 5.28/264\\ 
\hline
Binary Outcome (COVID Status) &Average &~ \\
~&Classification Error (\%)&~ \\
\hline 
    RandMVLearn & 9.76 (0.74) & 170.26/1015, 48.47/72, 129.98/5800, 114.9/264\\ 
    SIDA & 6.86 (3.61) & 10.14/1015, 11.8/72, 61.84/5800, 29.52/264\\ 
    Deep GCCA + SVM & 18.08 (1.23) & - \\ 
    MLP & 10.82 (4.40) & -  \\ 
\hline
\end{tabular}
\end{small}
\end{center}
\end{table}

\section{Discussion}
\label{s:Disc}

We have developed scalable kernel-based nonlinear methods to jointly integrate data from multiple sources and predict a clinical outcome. Our framework assumes that there is a shared or view-independent low-dimensional representation of all the views and that it can be estimated from nonlinear functions of the views. We used kernel methods to model the nonlinear functions and we restricted these functions to reside in a reproducing kernel Hilbert space. 
Based on the idea that random Fourier bases can approximate shift-invariant kernel functions, we constructed nonlinear mappings of each view and we used these nonlinear mappings and the outcome variable to learn  view-independent low-dimensional representations. As a result, the proposed methods can reveal nonlinear patterns in the views and can scale to a large training size. When estimating these low-dimensional view-independent representations, an outcome variable is used, providing them with interpretation capabilities.  Through the randomized nonlinear mappings, we learn the variables that likely drive the underlying shared low-dimensional structure in the views.  Our method allows the integration of prior biological information about variables and improves the interpretability of our findings, setting it apart from other nonlinear methods for multiview learning and prediction.  We have developed a user-friendly algorithm in Python 3, specifically Pytorch, and interfaced it with R to increase the reach of our algorithm.  Through simulation studies, we showed that the proposed method outperforms several
other linear and nonlinear methods for multiview data integration. When the proposed methods were applied to gene expression, metabolomics, proteomics, and lipidomics
data pertaining to COVID-19, we identified several molecular signatures for severe
disease and COVID-19 status.  The pathways enriched in our candidate signatures for the severity and status of COVID-19 included those related to inflammation and immune function. 

Our proposed methods have some limitations. First, the methods assume that the variable groups do not overlap. If the groups overlap, it may be possible to reconstruct nonoverlapping groups. For this purpose, one can consider variable-variable connections in each group and then keep a variable in the group with many connections. 
This assumes that a variable with many connections in a certain group is likely to play a key role in that group compared to another group where it only has fewer  connections. Second, the proposed methods only model shared low-dimensional representations among the views. Future work may consider modeling both shared and view-dependent low-dimensional representations. 

In summary, we have developed scalable randomized kernel-based methods for jointly learning nonlinear relationships in multiview data and predicting a clinical outcome. The proposed methods can identify variables or groups of variables that have the potential to contribute to the nonlinear association of the views and the variation in the outcome. Despite the above limitations, we find the simulations and real data applications encouraging and believe that the proposed methods will motivate future applications. 

\section*{Funding and Acknowledgments}
The project described was supported by the Award Number 1R35GM142695 of the National Institute of General Medical Sciences of the National Institutes of Health. The content is solely the responsibility of the authors and does not represent the official views of the National Institutes of Health.

\section*{Data Availability Statement}
The data used were obtained from \cite{overmyer:2020}. 
We provide a Python code interfaced with R, \textit{RandMVLearn}, to facilitate the use of our method. Its source codes, along with a README file, will be made available at: \url{https://github.com/lasandrall/RandMVLearn}.

\bibliographystyle{unsrtnat}
\bibliography{randmvbio}  

\clearpage
\section*{Supplementary Materials}
\section*{More on Algorithm}
\subsection*{Group Variable Selection}
The optimization problem for the group variable selection is:
\begin{eqnarray} \label{eqn:uncons}
    \min_{\bgamma^d}f(\bgamma^d)= \mathcal{L}({\bgamma^d})  + \mathcal{P}^{\eta_1}_{\eta_2}(\bgamma^d)
\end{eqnarray}
where $\mathcal{L}({\bgamma^d})= \frac{1}{2n}\|\bG - \bZ^d\bA^d\|^2_F$ and $\mathcal{P}^{\eta_1}_{\eta_2}(\bgamma^d) = \eta_1\|\bgamma^d\|_{1} + \eta_2 \sum_{l=1}^{G^d}\|\sqrt{p_l^d}\bgamma_l^d\|_{2}$, $\eta_1=\eta\rho$, and $\eta_2=(1-\eta)\rho$.
Following ideas in \cite{beck2009fast}, we construct the quadratic function
$Q_L(\bgamma^d,\widetilde{\bgamma}^d) = \mathcal{L}({\bgamma^d}) + \langle{ \bgamma^d - \widetilde{\bgamma}^d, \nabla{\mathcal{L}(\widetilde{\bgamma}^d}}\rangle +  \mathcal{P}^{\eta_1}_{\eta_2}( \bgamma^d) +  \frac{L}{2}\| \bgamma^d - \widetilde{\bgamma}^d\|^2_2$, for an appropriately chosen $L >0$,
and use it to approximate $f(\gamma^d)$ at the point $\widetilde{\bgamma}^d$. The step size $L$ is chosen such that $f(\bgamma^d) \le Q_L(\bgamma^d,\widetilde{\bgamma}^d)$. 
A key step in FISTA implementation is the computation of the proximal operator (or map) for the penalty $\mathcal{P}^{\eta_1}_{\eta_2}(\bgamma^d)$. We follow ideas in \cite{yuan2011efficient} and \cite{sra2012fast} for the proximal operator. For completeness sake, we describe this below. The proximal operator for the sparse group lasso penalty $\mathcal{P}^{\eta_1}_{\eta_2}(\bgamma^d)$, for a given $\widehat{\bgamma}^d$, is defined as: 
\begin{eqnarray} \label{eqn:prox}
\pi^{\eta_1}_{\eta_2}(\widehat{\bgamma}^d)=\argmin_{{\bgamma}^d} \frac{1}{2}\|\widehat{\bgamma}^d - \bgamma^d\|^2_2 +   \mathcal{P}^{\eta_1}_{\eta_2}(\bgamma^d).
\end{eqnarray}
Let $\widehat{\bgamma}^d = \widetilde{\bgamma}- \frac{1}{L}\nabla{\mathcal{L}}(\widetilde{\bgamma}^{d})$ where  $\widetilde{\bgamma}$ is an affine combination of previous two iteration solutions. Then the solution $\bgamma^{d}= \argmin_{\widehat{\bgamma}}Q_L(\bgamma^d,\widetilde{\bgamma}^{d})$ minimizes  $\pi^{\eta_1}_{\eta_2}(\widehat{\bgamma}^d))$ \citep{yuan2011efficient}. The proximal operator for sparse group lasso can be obtained by finding the proximal operator for lasso and using it in the proximal operator for group lasso \cite{yuan2011efficient}. The proximal operator for lasso reduces to solving problem (\ref{eqn:prox}) with $\eta_1 > 0$, and $\eta_2=0$, which corresponds to setting $\eta=1$ for $\rho >0$; i.e. $\argmin_{{\bgamma}^d} \frac{1}{2}\|\widehat{\bgamma}^d - \bgamma^d\|^2_2 +  \rho\|\bgamma^d\|_{1}$. The solution to this problem is given by: $\gamma^d_{Lasso}= \mbox{sgn}(\hat{\bgamma}^d) \cdot\max(|\hat{\bgamma}^d| - \rho, 0)$. By Theorem 1 in \cite{yuan2011efficient}, the solution to the optimization problem $\pi^{\eta_1}_{\eta_2}(\widehat{\bgamma}^d) = \pi^{0}_{\eta_2}(\bgamma^d_{Lasso})$. Therefore, for the sparse group lasso solution, we solve the proximal operator problem: 
\begin{eqnarray} \label{eqn:proxgroup}
\pi^{0}_{\eta_2}(\bgamma^d_{Lasso})=\argmin_{{\bgamma}^d}  \frac{1}{2}\|{\bgamma}^d - \bgamma^d_{Lasso}\|^2_2 +   \eta_2 \sum_{l=1}^{G^d}\sqrt{p_l^d}\|{\bgamma}_l^d\|_{2}.
\end{eqnarray}
By Lemma 2 in \cite{yuan2011efficient}, if the $l$th group satisfies $\|\bgamma^d_{Lasso_l}\|_2 \le \eta_2\sqrt{p_l^d}$, then the $l$th group has zero solution, i.e. $\widehat{\bgamma}^d_{l} =\mathbf{0}$. We cycle through for all groups and obtain the zero groups. We solve the optimization (\ref{eqn:proxgroup}) for the remaining groups, i.e. the nonzero groups. Since the groups are independent, we optimize (\ref{eqn:proxgroup}) for each  group: 
$ \pi^{0}_{\eta_2}(\bgamma^d_{Lasso_l})=\argmin_{{\bgamma}^d_l} \frac{1}{2}\|{\bgamma}^d_l - \bgamma^d_{Lasso_l}\|^2_2 +   \eta_2 \sqrt{p_l^d}\|\bgamma_l^d\|_{2}$, and  the solution for the $l$th group at that iteration is  given by $\widehat{\bgamma}^d_{SGLasso_l} = \max(1-\eta_2 \sqrt{p_l^d}\|\bgamma^d_{Lasso_l}\|_2^{-1},0)\bgamma^d_{Lasso_l}$\cite{sra2012fast}. 

To summarize, at iteration $t+1$, we obtain the gradient of the loss function $\mathcal{L}({\bgamma^d})$ with respect to $\gamma^d$ (i.e. $\nabla{\mathcal{L}}(\bgamma^{d}$)). We compute the gradient descent step: $\widetilde{\bgamma}^d=\bgamma^d - (1/L)\nabla{\mathcal{L}}(\bgamma^{d})$, where $L$ is the step size. We solve the proximal operator problem $\pi^{\eta_1}_{\eta_2}(\widetilde{\bgamma})$ given in equation (\ref{eqn:prox}); that is we obtain the lasso solution $\bgamma^d_{Lasso}$.  For the $l$th group, we check  if the condition $\|\bgamma^d_{Lasso_l}\|_2 \le \eta_2\sqrt{p_l^d}$ is satisfied. If it is, we set the $l$th group to zero, i.e. our group lasso solution for that group is zero,  $\widehat{\bgamma}^d_{SGLasso_l}=\mathbf{0}$.  If not, we obtain the solution $\widehat{\bgamma}^d_{SGLasso_l}$. We cycle through for all $l=1,\ldots,G$ groups. We update the step size $L$  according to the Armijo-Goldstein rule  $f(\widehat{\bgamma}^i_{SGLasso}) \le Q_L(\widehat{\bgamma}^i_{SGLasso},\widetilde{\bgamma}^i)$. We update  $\widetilde{\bgamma}^i$ using the previous two iterations,  and we iterate the process until convergence or until some maximum number of iterations is reached. We summarize our optimization process in  Algorithm \ref{alg::sparsegroupPGD}.



\subsection*{Individual Variable Selection}
We consider the following unconstrained optimization problem to learn $\bgamma^d$ for individual variable selection: \begin{eqnarray} \label{eqn:unconsInd}
    \min_{\bgamma^d}f(\bgamma^d)= \mathcal{L}({\bgamma^d})  + \mathbf{1}^{\smt}\bgamma^d
\end{eqnarray}
where $\mathbf{1}$ is a $p^d$-length vector of ones and $0 \le \bgamma^d \le 1$. We implement an accelerated projected gradient descent algorithm with backtracking. We use the probability simplex algorithm proposed in \cite{wang2013projection}  in the projection step of our gradient descent algorithm (Algorithm \ref{alg::PGD})
\subsection*{Hyperparameter Selection}
Our proposed optimization in equation (9) has several hyperparameters that deserve further discussion: $\lambda^d$ balancing model fit and complexity; number of latent components $r$; kernel parameters $\nu$ for the Gaussian kernel; size of random features, $M$; $\eta$ and $\rho$ for sparse group variable selection. We found $\lambda^d$ to be insensitive, so we set this at $1$ for all views. Regarding the size of the random feature, setting $M=300$ in our simulations with a sample size range of $500$ to $5,000$ resulted in competitive or superior prediction and/or variable selection performance. In our real data application where the sample size was $120$, we obtained competitive prediction estimates when we set $M=45$. Based on our simulations and the application of real data, we recommend setting $M=300$ for a sample size larger than $1,000$ and $M \approx \frac{n}{2}$ for a sample size smaller than $1,000$. 
However, we encourage exploring other random feature sizes. As for the Guassian kernel parameter (or bandwidth), $\nu$, if $\nu \rightarrow 0$, the Gram matrix becomes the identity matrix; if $\nu \rightarrow \infty$, all entries in the Gram matrix approach 1. In both cases, we lose all relevant information about the data.  We use median heuristic \citep{garreau2017large} to choose $\nu$ for each view. We choose  the number of latent components as follows. For each view, we obtain eigenvalues, $\lambda_1 > \lambda_2 \cdots > \lambda_r \cdots > \lambda_n$ from the Gram matrix  $\mathcal{K}^d(\mathbf{X},\mathbf{X})$ (few  eigenvalues could be obtained in situations where the samples size is large). Then for a component $r>2$, we compare the change in eigenvalue from the previous component, $r-1$ (i.e. $\frac{\lambda_r - \lambda_{r-1}}{\lambda_r}$). We choose the $r$ that results in a proportion of change less than a threshold, say $0.1$ for that view. The total number of latent components, $r$ is then set to the minimum number of components in the view-specific number of components. We recognize that there are sophisticated ways to select the number of components, but we prefer this simple approach due to its low computational burden and given that it yielded comparable or better results. The optimal hyperparameters $\eta$ and $\rho$ for sparse group lasso can be searched from a two-dimensional grid of $\eta$ and $\rho$, but this may be computationally expensive. Therefore, we fix the mixing parameter $\eta$ and obtain solutions for different values of $\rho$ since $\rho$ controls the amount of sparsity. The hyperparameter search space for $\rho$ could be large if many views have prior group information. To reduce computations,  we obtain all combinations of $\rho$ and randomly select some combinations to search from \citep{bergstra2012random}. 

\subsection*{Initializations for Algorithms}
For individual variable selection,  we initialize $\gamma^d$ evenly on a probability simplex. For group selection, we further weight the groups in $\gamma^d$ by their group weights.  We randomly sample $\epsilon^d_j,j=1,\ldots,M$ from the inverse Fourier of a shift-invariant Gaussian kernel. We initialize $\bG$ as an $n \times r$  matrix such that $\bG^{\smt}\bG=\bI_r$. We initialize $\mathbf{\Theta}$ as $r \times K-1$ (for categorical outcomes) random matrix or $r \times q$ (for multiple continuous outcomes) matrix of zeros. We initialize $\btheta$ as a length-$r$ vector of zeros for a single continuous outcome.  We initialize $\mathbf{A}^d$ as an $M \times r$ random matrix and we normalize so that $\|\bA^d\|_F=1$.

\begin{small}
\setlength{\intextsep}{2pt}
\begin{algorithm}[H]
	\DontPrintSemicolon
	\SetKwComment{tcp}{$\triangleright\ $}{}
	\textbf{Input}: Training data $(\bX^1,\ldots,\bX^D,\mathbf{y})$; tuning parameters $\lambda^d$, $M$ (number of random features), $d=1,\ldots,D$; for group variable selection, $\rho$ and $\eta$; $T$, maximum number of iterations \\  \;
	\textbf{Output}: Estimated $\bG$, $\btheta$ (or $\mathbf{\Theta}$), $\bA^d$, $\bgamma^d, d-1,\ldots,D$ \\ \;
	\textbf{Initialize}: $\bG$,$\btheta$ (or $\mathbf{\Theta}$), $\bgamma^d,d=1,\ldots,D$ evenly on simplex $\Delta$; for group variable selection obtain $\bgamma^d_l/\sqrt{p^d_l}$;~~ $\epsilon^d_j \sim p(\bw)$, $b^d_j \sim U[0, 2\pi]$; 
	\text{rescalings}: $\mathbf{\omega}^d_j= \bgamma^d \cdot \epsilon^d_j, \forall j =1,\ldots,M; d=1,\ldots,D$; construct random features $\mathbf{Z}^d$ \\\;
    \textbf{Objective}: 
    \begin{footnotesize}
$\mathcal{L}(\mathbf{Y}, \bG, \mathbf{\Theta}) +  \frac{1}{2n}\sum_{d=1}^{D}\|\bG - \bZ^d\bA^d\|^2_F +\frac{\lambda^d}{2} \sum_{d=1}^{D}\|\bA^d\|^2_F  $\;
    \end{footnotesize}
		\tcp*{{\small  $\bZ^d$ depends on $\bgamma^d$}}
     \Repeat{ convergence or maximum number of iterations $T$ reached}{
		\Begin{ Step 1: Solve for $\bgamma^d$\;
 \For{$d=1,\ldots,D$}{ 	
  $\bgamma^d \leftarrow   
    \min_{\bgamma^d}\frac{1}{2n}\|\bG - \bZ^d\bA^d\|^2_F +\mathcal{P}(\bgamma^d)$\;
\tcp*{{\small  $\bZ^d$ is a function of $\gamma^d$. Call Algorithm 2 or 3. }}\;
 Rescalings:  $\mathbf{\omega}^d_j= \bgamma^d \cdot \epsilon^d_j, \forall j =1,\ldots,M$\;
 Construct random features $\mathbf{Z}^d$ 
}
  		}
	
	\Begin{ Step 2: Solve for $\bA^d$, $\bG$, $\btheta$ (or $\mathbf{\Theta}$)  \;
       \For{$d=1,\ldots,D$}{
       $\bA^d \leftarrow     \min_{\bA^d}\frac{1}{2n} \|\bG - \bZ^d\bA^d\|^2_F + \frac{\lambda^d}{2} \|\bA^d\|^2_F$\;
       }
 	\tcp*{{\small  least squares solution}}      

 Solve  $ \mathbf{G} \leftarrow  \min_{\bG:\bG^{\smt}\bG=\bI_r} \left\{\mathcal{L}(\mathbf{Y}, \bG, \mathbf{\Theta}) +  \frac{1}{2n}\sum_{d=1}^{D}\|\bG - \bZ^d\bA^d\|^2_F  \right\}$\;
\tcp*{{\small  Procrustes problem}}\;
 Solve  $\mathbf{\Theta} \leftarrow     \min_{\mathbf{\Theta}}\mathcal{L}(\mathbf{Y}, \bG, \mathbf{\Theta})$\;
 \tcp*{{\small closed form solution}}\;
   		}

   		}
\caption{Algorithm for Scalable Randomized Kernel Methods for Multiview Learning}
\label{alg::main}
\end{algorithm}
\end{small}
\clearpage
\setlength{\intextsep}{2pt}
\begin{algorithm}[H]
	\DontPrintSemicolon
	\SetKwComment{tcp}{$\triangleright\ $}{}
	\textbf{Input}: $\bG$, $\bA^d$, random features $\bZ^d$,  $\bgamma^d$, $T$ (maximum iteration); $G^d$ (number of groups for view $d$); $\rho$, $\eta$ \\  \;
	\textbf{Output}: Estimated $\bgamma^d$  \\  \;
    \textbf{Objective}: 
      $f(\bgamma^d)=\frac{1}{2n}\|\bG - \bZ^d\bA^d\|^2_F + \mathcal{P}^{\eta_1}_{\eta_2}(\bgamma^d)$\;
		\tcp*{{\small  $\bZ^d$ depends on $\bgamma^d$}. We suppress the superscript $d$ in $\bgamma^d$ for ease of notation and only write iteration number)}
	\textbf{Initialize (Step 0)}: $L>0$, $L^0=L$, $\beta^{-1}=0$, $\beta^{0}=0$, $\bgamma^0=\bgamma^d$ \;
	
   \Repeat{ convergence or maximum iteration $T$}{
     \textbf{Step $t (t \ge 1)$}  \;
     Set $\alpha^t= \frac{\beta^{t-2}-1}{\beta^{t-1}}$; $\widetilde{\bgamma}^{t}=\bgamma^t + \alpha^t(\bgamma^t-\bgamma^{t-1})$ \\ \;
     Find the smallest $L=2^sL^{t-1}$, $s=0,1,\ldots,$ such that $f(\widehat{\bgamma}^t_{SGLasso}) \le Q_L(\widehat{\bgamma}^t_{SGLasso},\widetilde{\bgamma}^t)$ where $\widehat{\bgamma}^t_{SGLasso}$ is obtained as follows.\;
     Find gradient of loss  and construct descent step: $\widehat{\bgamma}^{t} = \widetilde{\bgamma}^{t}- \frac{1}{L}\nabla{\mathcal{L}}(\widetilde{\bgamma}^t)$ \\ \;
	 \tcp*{{\small  $\mathcal{L}(\bgamma^t)= \frac{1}{2n}\|\bG - \bZ^d\bA^d\|^2_F$. $\bZ^d$ depends on $\bgamma$}}
	  Find the lasso solution:  $\widehat{\bgamma}^{t}_{Lasso}=\mbox{sgn}(\widehat{\bgamma}^{t}) \cdot\max(|\widehat{\bgamma}^{t}| - \rho, 0)$ \\ \;
	 {
	 \textbf{for}  $l=1,\ldots,G^d$ \textbf{do} \\ \;
	 ~~~~~~~If $\|\widehat{\bgamma}^t_{Lasso_l}\|_2 \le \eta_2\sqrt{p_l^d}$, then $\widehat{\bgamma}^i_{SGLasso_l} =\mathbf{0}$ \\ \; 
	 ~~~~~~~If not, then  $\widehat{\bgamma}^t_{SGLasso_l}= \max(1-\eta_2 \sqrt{p_l^d}\|\widehat{\bgamma}^t_{Lasso_l}\|_2^{-1},0)\widehat{\bgamma}^t_{Lasso_l}$ \\ \;
	 \textbf{end for}\\  \;
	 Set: ~~~~~$L^t=L$ \\
	 ~~~~~~~~~~~${\bgamma}^{t} =\widehat{\bgamma}^t_{SGLasso}$\\
		~~~~~~~~~~~~$\beta^{t+1}=\frac{1+ \sqrt{1 + 4(\beta^{t})^2}}{2}$\\
   		}
   		}
   		  		Set $\gamma^d =\bgamma^t$ (Sparse Group Lasso Solution for View $d$)
\caption{Fast Sparse Group Lasso  for View $d$}
\label{alg::sparsegroupPGD}
\end{algorithm}

\clearpage
\setlength{\intextsep}{2pt}
\begin{algorithm}[H]
	\DontPrintSemicolon
	\SetKwComment{tcp}{$\triangleright\ $}{}
	\textbf{Input}: $\bG$, $\bA^d$, random features $\bZ^d$, $\bgamma^d$, $T$ (maximum iteration) \;
	\textbf{Output}: Estimated $\bgamma^d$  \;
    \textbf{Objective}: 
      $f(\bgamma^d)=\frac{1}{2n}\|\bG - \bZ^d\bA^d\|^2_F + \mathbf{1}^{\smt}\bgamma^d$\;
		\tcp*{{\small  $\bZ^d$ depends on $\bgamma^d$}. We suppress the superscript $d$ in $\bgamma^d$ for ease of notation and only write iteration number)}
	\textbf{Initialize (Step 0)}: $L>0$, $L^0=L$, $\beta^{-1}=0$, $\beta^{0}=0$, $\bgamma^0=\bgamma^d$ \;
	
   \Repeat{ convergence or maximum iteration $T$}{
     \textbf{Step $t (t \ge 1)$}  \;
     Set $\alpha^t= \frac{\beta^{t-2}-1}{\beta^{t-1}}$; $\widetilde{\bgamma}^{t}=\bgamma^t + \alpha^t(\bgamma^t-\bgamma^{t-1})$\;
     Find the smallest $L=2^sL^{t-1}$, $s=0,1,\ldots,$ such that $f(\widehat{\bgamma}^t_{Ind}) \le Q_L(\widehat{\bgamma}^t_{Ind},\widetilde{\bgamma}^t)$ where $\widehat{\bgamma}^t_{Ind}$ is obtained as follows.\;
	 {Find gradient of loss  and construct descent step: $\widehat{\bgamma}^{t} = \widetilde{\bgamma}^{t}- \frac{1}{L}\nabla{\mathcal{L}}(\widetilde{\bgamma}^t)$\;
	 \tcp*{{\small  $\mathcal{L}(\bgamma^t)= \frac{1}{2n}\|\bG - \bZ^d\bA^d\|^2_F$. $\bZ^d$ depends on $\bgamma$}}
	 Projection Step: $\bgamma^{t}_{Ind}= \min_{\bgamma^d \in \Delta}\frac{1}{2}\|\bgamma^t - \widehat{\bgamma}^{t} \|^2_2$\;
	 \tcp*{{Use algorithm in \cite{wang2013projection}}}
	 Set: ~~~~~$L^t=L$ \\
	  ~~~~~~~~~~~${\bgamma}^{t} =\widehat{\bgamma}^t_{Ind}$\\
		~~~~~~~~~~~~$\beta^{t+1}=\frac{1+ \sqrt{1 + 4(\beta^{t})^2}}{2}$\\
   		}
   		}
   		  		Set $\gamma^d =\bgamma^t$ (Individual variable sparsity for view $d$)
\caption{ Fast Individual Variable Selection for View $d$}
\label{alg::PGD}
\end{algorithm}

\clearpage
\section*{More on Simulations}
\subsection*{Continuous Outcome}
\begin{figure}[htb!]
\begin{tabular}{ccc}
         \centering
         \includegraphics[width=0.3\textwidth]{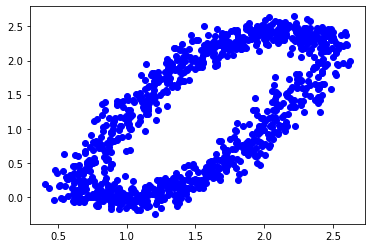} & \includegraphics[width=0.32\textwidth]{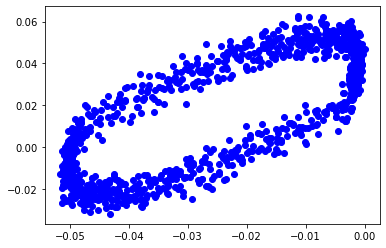}&\includegraphics[width=0.37\textwidth]{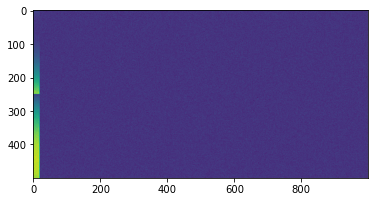}\\
\end{tabular}
    \caption{Top Panel (Left): Relationship between signal variables 1 and 2 in View 1 ; (Middle): plot of first two common low-dimensional representation between Views 1 and 2 (i.e., G). (Right): Image plot of View 1. First twenty variables are signals and remaining variables are noise. Plots of View 2 are similar to View 1. }
    \label{fig:nonlinearcont}
\end{figure}

We consider four different settings that differ in their number of samples and variables. In each setting,  20 variables have nonlinear relationships (Figure \ref{fig:nonlinearcont}) and are considered signal variables. We generate data for the two Views, $\mathbf{X}^1$ and $\mathbf{X}^2$, and the outcome $\mathbf{y}$ as follows.
\begin{itemize}
    \item Generate $\btheta$ as a vector of $n/2$ evenly spaced points between $0.6$ and $2.5$
    \item Form the vector $\mathbf{s}=[(\btheta -\mathbf{1})^2 ;~ (\btheta +0.1\mathbf{1})^2 -2 (\btheta -\mathbf{1})^2] \in \Re^{n \times 1}$
    \item Set $\widetilde{\bX}_1 =[\tilde{\btheta}; \mathbf{s}\mathbf{1}^{\smt}_{p^1-1} ] \in \Re^{n \times p^1}, \tilde{\btheta}=[\btheta;\btheta] \in \Re^{n \times 1}$ 
    \item Generate $\mathbf{X}^1= \widetilde{\bX}_1 \cdot \bW + \sigma_{1}\bf{E}^1$ where $(\cdot)$ is element-wise multiplication,  $\bW \in \Re^{n \times p^1}= [\mathbf{1}_{20}, \mathbf{0}_{p^1-20}]$ is a matrix of ones and zeros, $\mathbf{1}$ is a matrix of ones, $\mathbf{0}$ is matrix of zeros, $\sigma_1 =0.1$, and $\mathbf{E}^1 \sim N(0,1)$. By generating $\mathbf{X}^1$ this way, we assume the first $20$ variables have nonlinear relationships and are informative.  
    \item Generate the second view as:  ${\bX}^2=5\mathbf{X}^1 + \sigma_{2}\mathbf{E}^2$, where $\mathbf{E}^2 \sim N(0,1)$, and $\sigma_2=0.2$.
    \item Generate the continuous outcome, $\mathbf{y}$ as $\mathbf{y}=5\mathbf{G}\btheta + \sigma_{y}\mathbf{e}_{y}$, where the columns in $\mathbf{G} \in \Re^{n \times 3}$ are the first three left singular vectors from the singular value decomposition of the concatenated data $[\mathbf{X}_1, \mathbf{X}_2]$, $\btheta \sim U(0,1) \in \Re^{2 \times 1}$, $\sigma_{y}=0.3$, and $\mathbf{e}_y \sim N(0,1)$. 
\end{itemize}
\subsection*{Continuous Outcome}
We compare the proposed methods, RandMVLearn and RandMVLearnG (for the selection of group variables), with linear and nonlinear methods for associating data from multiple views. For linear methods, we consider the following methods: sparse canonical correlation analysis [sCCA] \citep{safo2018sparse}, canonical variate regression (CVR) \cite{CCAReg2016} and sparse integrative discriminant analysis for multiview data (SIDA) \cite{SIDA:2019}. For nonlinear methods, we compare our methods with the following methods: deep canonical correlation analysis (Deep CCA) \citep{Andrew:2013} and MOMA \cite{MOMA:2022}. SIDA is a one-step method for simultaneously assessing associations among multiple views and classifying a subject into one of two or more classes. Since sCCA and Deep CCA are mainly used for associating multiple views, we use the scores from these methods in linear or nonlinear regression models to investigate the prediction and classification performance of the low-dimensional representations learned from these methods. To this end, we concatenate the canonical variates from both views. To examine the advantages of appropriately integrating data from multiple sources, these integrative analysis methods are compared to naive implementations of    multiple layer perceptron (MLP)-- for continuous outcome-- and support vector machine (SVM) \citep{ben2008support} --for binary outcome-- on stacked views. We perform sCCA and SIDA with the respective \textit{SelpCCA} and SIDA R packages  provided by the authors on Github. We set the number of random features in our proposed methods at 300. We use median heuristic to select the bandwidth of the Gaussian kernel for each view. 
We perform Deep CCA using PyTorch codes provided by the authors. To rank variables, we combine Deep CCA with the Teacher-Student Framework (TS) \citep{TS:2019}. We compare the TS feature selection approach to the proposed methods.  We perform MOMA using Python code provided by authors. 

We compare the performance of the proposed variable selection options- individual and group variable selections.
For group variable selection, there are two groups of variables: a group of signal variables and a group of noise variables.  We fix $\eta=0.5$ in the sparse group lasso penalty. We consider $5$ values of $\rho$ in the range $(0, \rho_{max})$ for each view. $\rho_{max}$ was chosen to prevent trivial solutions in $\gamma^d_{SGLasso}$. We perform a grid and random search for plausible combinations of $\rho$ from each view, and choose the optimal $\rho$ combination via three-fold cross-validation.  

\begin{figure}[htb!]\label{fig:cont}
\centering
\includegraphics[scale=0.5]{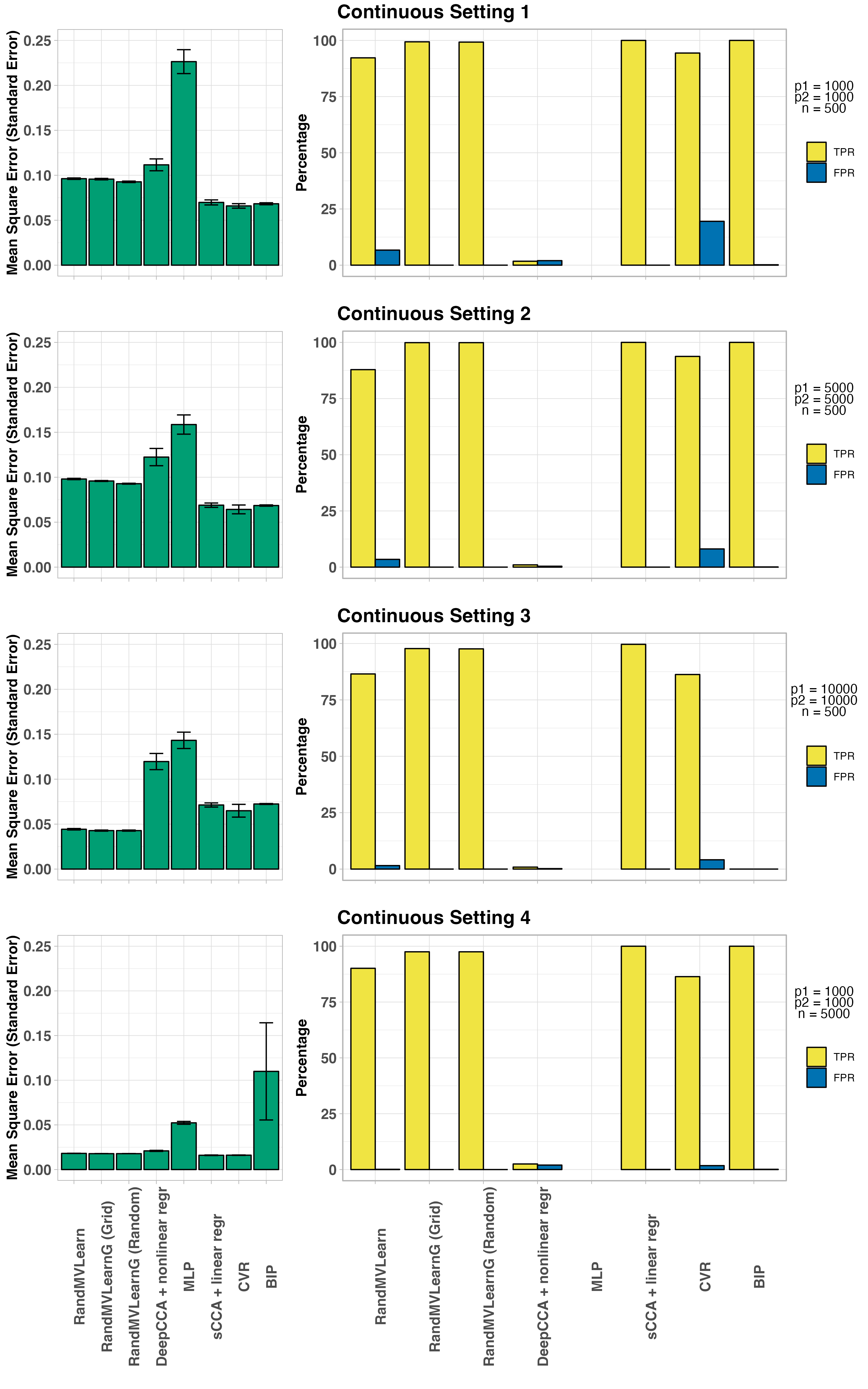}
\caption{Simulation results for continuous settings. Number of random features, $M$ set to 300. Number of components fixed at $K=5$. The proposed methods RandMVLearn, RandMVLearnG have lower error rates in almost all settings, higher or comparable true positive rates (TPR), and lower or comparable false positive rates (FPR)}.
\end{figure}

Figure 2 gives the prediction and variable selection performance of the proposed and comparison methods. We combine the variable selection performance of the two views and report averages across the 20 repetitions. The MSEs for our approach with individual variable selection (RandMVLearn) are comparable to that of the group variable selection (i.e., RandMVLearnG). However, the variable selection performance of the group sparsity approach is better than that of the individual approach, highlighting the advantages of using group information when it is available. Compared to CVR, an approach for simultaneously modeling linear relationships among views and predicting an outcome, our proposed methods have better variable selection performance and competitive error rate. In particular, the MSEs for our proposed methods are lower when the number of variables is large and comparable when the number of samples is large. The variable selection performance of RandMVLearnG is comparable to that of sCCA in all settings and better than that of MOMA. In terms of MSE, sCCA followed by linear regression resulted in lower MSE's in all settings, except in Setting 3.

\section*{More on Real Data Analysis}
\subsection*{Prediction and Variable Selection }
We consider the sparse  group implementation of the proposed method assuming that group information exists for the RNA-Seq data. For this purpose, we followed ideas in \cite{simon2013sparse} and grouped the genes into ``genesets" using cytogenetic position data (the C1 set from \cite{subramanian2005gene}). Of the  5,800 genes,   5,716 were found in the C1 genesets, and we removed the others from the analysis. The C1 set contains 299 genesets corresponding to each human chromosome and each cytogenetic band. We had data for 291 genesets. We chose the C1 collection because it had no overlapping groups. For the remaining views, we used the individual variable selection approach. The analytical data for the RNA sequencing data  in the group variable selection application is $ \mathbf{X}^{3} \in \Re^{120 \times 5,716}$. data were randomly divided into 50 training sets ($n=71$) and testing sets ($n=49$)  while maintaining proportions of COVID-19 status similar to those of the complete data. We used the training data to fit the models.  We used the testing data to assess error rates-- MSE and misclassification rates for continuous and binary outcomes, respectively.  We set the number of random features to 45, to be smaller than the training and testing sizes. We chose the number of shared  low-dimensional representations using the simple approach proposed. We chose the kernel parameter using median heuristic. We fix $\eta=0.5$. For the sparse group implementation on the RNA-Seq data, we considered $8$ values of $\rho$ in the range $(0, \rho_{max})$. $\rho_{max}$ was chosen to prevent trivial solutions in $\gamma^d_{SGLasso}$. The optimal $\rho$ was chosen via three-fold cross-validation.  

We compare the prediction estimates of RandMVLearn with BIPNet \citep{chekouo2023bayesian}, a Bayesian method for simultaneous data integration and prediction that allows to incorporate group information. BIPNet learns linear relationships in the views and the outcome. For BIPNet, we set the marginal prosterior probability for selecting variables to 0.9.  From Table 1, the average prediction estimate of RandMVLearn is comparable to BIPNet. RandMVLearnG resulted in suboptimal prediction estimate. 

\begin{table}
\caption{Prediction Estimates and Variable Selection. L: Lipidomics; M:Metabolomics; R: RNA-Seq ; P: Proteomics; RandMVLearn is the proposed method without group information. MSE refers to mean square error}
\label{tab: covid19}
\begin{center}
\begin{tabular}{lrr}
\hline
Method&Average MSE (Std Error) & Average \# of variables selected	\\
~ & ~ & (L,M,R,P)\\
			\hline
			\hline
Continuous (Prior) & ~&~ \\
\hline
            RandMVLearnG  & 1.0029 (0.0183)& 76.08/1015, 33.9/72, 63.0/5716, 55.37/264 \\
            RandMVLearn & 0.8888 (0.0212) & 142.48/1015, 47.48/72, 207.86/5716, 110.60/264 \\
		BIPnet with MPP threshold 0.9 & 0.8796 (0.0165) & 179.20/1015, 7.52/72,  2932.82/5716, 5.96/264\\
			
\hline
\end{tabular}
\end{center}
\end{table}

\subsection*{Pathway, Disease and Functions Enriched in Protein List (No Group Information)}
\begin{table}[!h]
\caption{Top 15 IPA significantly (p-value $< 0.05$) enriched pathways in Protein list for COVID-19 status}
\label{tab: covid19statusPathway}
\begin{center}
\begin{tiny}
\begin{tabular}{lll}
\hline
Pathway name & -log(p-value) & molecules  	\\
\hline
\hline
p70S6K Signaling&	11.8&	AGT,IGHV1-24,IGHV3-20,IGHV3-21,IGHV3-33,\\
~&~&IGHV3-53,IGHV3-74,IGKV2-30,IGLV3-21,IGLV3-27,IGLV4-69,YWHAE	\\
B Cell Development&	11.2&	HLA-A,IGHV1-24,IGHV3-20,IGHV3-21,IGHV3-33,\\
~&~ &IGHV3-53,IGHV3-74,IGKV2-30,IGLV3-21,IGLV3-27,IGLV4-69	\\
IL-15 Signaling&	9.4&	IGHV1-24,IGHV3-20,IGHV3-21,IGHV3-33,\\
& & IGHV3-53,IGHV3-74,IGKV2-30,IGLV3-21,IGLV3-27,IGLV4-69	\\
Fc$\gamma$RIIB Signaling in B Lymphocytes&	9.36&	IGHV1-24,IGHV3-20,IGHV3-21,IGHV3-33,IGHV3-53,\\
& & IGHV3-74,IGKV2-30,IGLV3-21,IGLV3-27,IGLV4-69	\\
PI3K Signaling in B Lymphocytes&	8.93&	IGHV1-24,IGHV3-20,IGHV3-21,IGHV3-33,IGHV3-53,\\
& & IGHV3-74,IGKV2-30,IGLV3-21,IGLV3-27,IGLV4-69	\\
 B Cell Receptor Signaling&	8.63&	IGHV1-24,IGHV3-20,IGHV3-21,IGHV3-33,IGHV3-53,\\
Communication between Innate and Adaptive Immune Cells&8.2&	HLA-A,IGHV1-24,IGHV3-20,IGHV3-21,IGHV3-33,IGHV3-53,\\
& & IGHV3-74,IGKV2-30,IGLV3-21,IGLV3-27,IGLV4-69	\\
Altered T Cell and B Cell Signaling in Rheumatoid Arthritis&	8.17&	HLA-A,IGHV1-24,IGHV3-20,IGHV3-21,IGHV3-33,IGHV3-53,\\
& & IGHV3-74,IGKV2-30,IGLV3-21,IGLV3-27,IGLV4-69	\\
Systemic Lupus Erythematosus Signaling&	8.08&	HLA-A,IGHV1-24,IGHV3-20,IGHV3-21,IGHV3-33,IGHV3-53,\\
& & IGHV3-74,IGKV2-30,IGLV3-21,IGLV3-27,IGLV4-69	\\
Role of NFAT in Regulation of the Immune Response&	7.73&	HLA-A,IGHV1-24,IGHV3-20,IGHV3-21,IGHV3-33,IGHV3-53,\\
& & IGHV3-74,IGKV2-30,IGLV3-21,IGLV3-27,IGLV4-69	\\
Systemic Lupus Erythematosus In B Cell Signaling Pathway&	7.6&	IGHV1-24,IGHV3-20,IGHV3-21,IGHV3-33,IGHV3-53,\\
 & & IGHV3-74,IGKV2-30,IGLV3-21,IGLV3-27,IGLV4-69	\\
Phospholipase C Signaling&	6.31&	IGHV1-24,IGHV3-20,IGHV3-21,IGHV3-33,IGHV3-53,IGHV3-74,\\
& & IGKV2-30,IGLV3-21,IGLV3-27,IGLV4-69	\\
LXR/RXR Activation&	4.82&	AGT,APOA4,LPA,S100A8	\\
Atherosclerosis Signaling&	4.69&	APOA4,ICAM1,LPA,S100A8	\\
Role of IL-17A in Psoriasis&	3.88&	S100A8,S100A9	\\

\hline
\end{tabular}
\end{tiny}
\end{center}
\end{table}

\begin{table}[!h]
\caption{Top 15 IPA significantly (p-value $< 0.05$) enriched pathways in protein list for COVID-19 severity}
\label{tab: covid19statusPathway}
\begin{center}
\begin{tiny}
\begin{tabular}{lll}
\hline
Pathway name & -log(p-value) & molecules  	\\
\hline
\hline
p70S6K Signaling&	10.4&	AGT,IGHV1-18,IGHV1-24,IGHV1-3,IGHV2-70,IGKV1-17,\\
& & IGKV2-30,IGLV3-19,IGLV4-69,YWHAE\\								
B Cell Development&	8.1&	IGHV1-18,IGHV1-24,IGHV1-3,IGHV2-70,IGKV1-17,\\
& & IGKV2-30,IGLV3-19,IGLV4-69	\\							
IL-15 Signaling&	7.84	&IGHV1-18,IGHV1-24,IGHV1-3,IGHV2-70,IGKV1-17,\\
& & IGKV2-30,IGLV3-19,IGLV4-69\\								
Fc$\gamma$RIIB Signaling in B Lymphocytes	&7.81&	IGHV1-18,IGHV1-24,IGHV1-3,IGHV2-70,\\
& & IGKV1-17,IGKV2-30,IGLV3-19,IGLV4-69					\\			
PI3K Signaling in B Lymphocytes	&7.46&	IGHV1-18,IGHV1-24,IGHV1-3,IGHV2-70,IGKV1-17,\\
& & IGKV2-30,IGLV3-19,IGLV4-69				\\				
B Cell Receptor Signaling	&7.22	&IGHV1-18,IGHV1-24,IGHV1-3,IGHV2-70,IGKV1-17,\\
& & IGKV2-30,IGLV3-19,IGLV4-69\\	
LXR/RXR Activation	&7.06&	AGT,APOC4,LBP,LPA,S100A8\\	
Systemic Lupus Erythematosus In B Cell Signaling Pathway	&6.78&	IGHV1-18, IGHV1-24,IGHV1-3,\\
& & IGHV2-70,IGKV1-17,IGKV2-30,IGLV3-19,IGLV4-69	\\							
Communication between Innate and Adaptive Immune Cells&	5.95&	IGHV1-18,IGHV1-24,IGHV1-3,IGHV2-70,\\
& & IGKV1-17,IGKV2-30,IGLV3-19,IGLV4-69	\\							
Altered T Cell and B Cell Signaling in Rheumatoid Arthritis	&5.92&	IGHV1-18,IGHV1-24,IGHV1-3,IGHV2-70,\\
& & IGKV1-17,IGKV2-30,IGLV3-19,IGLV4-69\\								
Role of NFAT in Regulation of the Immune Response	&5.6&	IGHV1-18,IGHV1-24,IGHV1-3,IGHV2-70,IGKV1-17,\\
& & IGKV2-30,IGLV3-19,IGLV4-69		\\						
Systemic Lupus Erythematosus Signaling&	5.51&	IGHV1-18,IGHV1-24,IGHV1-3,IGHV2-70,IGKV1-17,\\
& & IGKV2-30,IGLV3-19,IGLV4-69		\\						
Phospholipase C Signaling	&5.35&	IGHV1-18,IGHV1-24,IGHV1-3,IGHV2-70,IGKV1-17,\\
& & IGKV2-30,IGLV3-19,IGLV4-69	\\							
Role of IL-17A in Psoriasis	&4.13&	S100A8,S100A9		\\						
Neuroprotective Role of THOP1 in Alzheimer's Disease	&3.72&	AGT,SERPINA3,YWHAE	\\							
\hline
\end{tabular}
\end{tiny}
\end{center}
\end{table}






\end{document}